\documentclass[aps,twocolumn,showpacs,superscriptaddress,prb,amsmath,amssymb,letterpaper]{revtex4}
\usepackage{times}
\usepackage{amsmath,bm,amsfonts}
\usepackage{dcolumn}
\usepackage{graphicx}
\usepackage{latexsym}

\begin{document}

\title{Searching for topological density wave insulators in  multi-orbital square lattice systems}

\author{Bohm-Jung \surname{Yang}}
\affiliation{Department of Physics, University of Toronto,
Toronto, Ontario M5S 1A7, Canada}

\author{Hae-Young \surname{Kee}}
\email[Electronic Address:]{hykee@physics.utoronto.ca}
\affiliation{Department of Physics, University of Toronto,
Toronto, Ontario M5S 1A7, Canada}
\affiliation{Canadian Institute for Advanced Research, Quantum Materials program, Toronto, Ontario M5G 1Z8, Canada}

\date{\today}

\begin{abstract}
We study topological properties of density wave states with broken translational
symmetry in two-dimensional multi-orbital systems with a particular focus
on t$_{2g}$ orbitals in square lattice.
Due to distinct symmetry properties of d-orbitals,
a nodal charge or spin density wave state with Dirac points protected by lattice symmetries
can be achieved. When an additional order parameter
with opposite reflection symmetry
is introduced to a nodal density wave state, the system can be fully gapped leading to a band insulator.
Among those, topological density wave (TDW) insulators can be realized, when
an effective staggered on-site potential generates a gap to
a pair of Dirac points connected by the inversion symmetry
which have the same topological winding numbers.
We also present a mean-field phase diagram for various density wave states, and discuss experimental implications
of our results.
\end{abstract}


\maketitle

\section{\label{sec:intro} Introduction}


Identifying topological insulators has been one of the most fascinating research fields
in contemporary condensed matter physics.~\cite{Qi10,Moore10,Hasan10}
Topological insulators have a bulk gap like band insulators, but are distinguished
by topologically protected conducting edge states preserving time-reversal invariance.
In particular, two-dimensional topological
insulators are known as quantum spin Hall insulators with finite counter-propagating spin currents
on the edge, analogous to quantum Hall states.
Haldane~\cite{Haldane} proposed that the fictitious magnetic fluxes in the honeycomb lattice lead to
the quantum anomalous Hall insulator (or Chern insulator).
Generalizing Haldane's model including time reversal invariant spin-orbit coupling,
it was theoretically shown that such a quantum spin Hall insulator can exist in graphene.~\cite{KaneMele2, Min}
A two-dimensional semiconductor system with a uniform strain gradient was also proposed to be a
candidate.~\cite{Zhang_QSH}
Later, the predicted edge states in HgCeTe quantum well systems~\cite{bernevig06} were
experimentally verified which confirmed the existence of two-dimensional topological insulators.~\cite{konig07}

The topological insulators in these systems normally exist due to strong spin-orbit coupling.~\cite{KaneMele2,KaneMele1}
When the spin-orbit coupling preserves spin rotational symmetry about an axis, the counter-propagating
edge modes which carry opposite spin quantum numbers result in quantum spin Hall insulators.
It was shown that these modes  are protected by time reversal symmetry even in the
absence of spin rotational invariance.~\cite{KaneMele1}
It was further pointed out that an effective spin-orbit coupling term can be generated by spontaneous
spin rotational symmetry breaking in an extended Hubbard model on the honeycomb lattice.~\cite{raghu08}
In these studies, the structure of the honeycomb lattice plays an important role, as the tight binding model
on this lattice possesses two Dirac points at the Brillouin zone corners.
Therefore in low energy description, various gapped insulating phases proximate
to the Dirac semi-metal can be understood in terms of mass perturbations
to gapless Dirac particles.
For instance, the fictitious magnetic fluxes introduced by Haldane
generate a mass term that has the opposite signs at the two Dirac points leading to
an insulator with finite quantized Hall conductivity.
The Dirac Hamiltonian approach further provides
a framework to understand the time-reversal invariant $Z_{2}$ topological insulators.~\cite{KaneMele1}

While systems on the honeycomb lattice such as graphene naturally support two-dimensional massless Dirac
particles in the bare band structures, this is not the case in a simple square lattice system which is an
effective model for abundant layered perovskite materials in nature.
In this respect, it is interesting to note that the recently proposed nodal density wave state~\cite{Ran}
exhibits gapless Dirac particles via broken translational symmetry.
This proposal was made in the context of iron pnictide systems,
where  d-orbitals of t$_{2g}$ bands in an effectively two-dimensional square lattice
give rise to several Fermi pockets with interesting topological properties.
In this system, the spin density wave instability
with the finite ordering wave vector $\textbf{Q}=(\pi,0)$ (or $(0,\pi)$)
leads to band touchings between the ${\bf k}$ states with the momentum difference of $\textbf{Q}$.
In general, the degeneracies at the band touching points disappear because of the finite overlap
matrix between the degenerate states induced by the density wave order parameter.
However, in multi-orbital systems, because of the distinct symmetry properties of orbitals,
the degeneracies at some band touching points are protected leading to nodal density wave states,
which is generally valid for any density wave orders.

In this work, we ask if topological insulators can be emerged by gapping nodal points turning
the system from nodal density wave states to topological density wave (TDW) insulators.
To find such a TDW insulator, we first investigate the properties of the
nodal density wave states.
We find that one general and important characteristic of the Dirac particles
in nodal density wave states is that
a pair of Dirac Hamiltonians connected by the inversion symmetry
have the same topological winding numbers.
Thus an effective staggered on-site potential generating a mass term,
which has the same signs at the inversion symmetric nodal points,
induces TDW insulators.
This can be contrasted with the topological properties
of the Dirac particles in the honeycomb lattice where
the Dirac Hamiltonians
at the two inversion symmetric nodal points have the opposite winding directions.~\cite{Graphene_review, Kane_review}
Thus the mass term induced by, for example, a staggered sublattice chemical potential,
which has the same signs at the two Dirac points would generate a topologically trivial
band insulator as shown in graphene system.\cite{KaneMele2,Semenoff}

The rest of the paper is organized as follows. In Sec. II,
we first consider a simple two-band model Hamiltonian
composed of $d_{xz}$ and $d_{yz}$ orbitals on the square lattice.
After classifying all possible charge and spin density
wave order parameters with the ordering wave vector $\textbf{Q}=(\pi,0)$
based on their transformation properties
under lattice symmetries, we establish general relations between
the locations of Dirac nodes and order parameter symmetries in Sec. III.
The fact that $d_{xz}$ and $d_{yz}$ orbitals have the opposite eigenvalues
under reflection symmetries along high symmetry directions
in the momentum space, plays the key role for the emergence of Dirac points.
In addition to the Dirac points coming from the Brillouin zone folding,
additional contributions from quadratic band degeneracy splitting are also discussed.
In Sec. IV, topological properties of gapped density wave phases
with two order parameters with the opposite reflection symmetries
are studied.
Fully-gapped insulating phases can be obtained by introducing
two density wave order parameters which have the opposite
eigenvalues under reflection symmetries.
Among them, a certain combination turns the system to a TDW insulator.
In Sec. V, the mean field phase diagram including
the  TDW phase is presented,
which is obtained by solving an extended Hubbard
model Hamiltonian with orbital degeneracy.
Topological density wave states in three orbital
systems are discussed in Sec. VI.
Straightforward extension to three-orbital systems
shows the general applicability of the idea we pursue in this work to obtain
topological insulators in multi-orbital systems.
Finally, we conclude in Sec. VII.

\section{\label{sec:meanfield} Two band Hamiltonian and symmetries of order-parameters}

\subsection{\label{sec:tightbinding} Tight-binding Hamiltonian}

We consider a tight binding Hamiltonian
on the square lattice with two orbital ($d_{xz}$, $d_{yz}$) degrees of freedom at each site.
A generic Hamiltonian which contains all possible hopping processes
allowed by lattice symmetries is given by

\begin{align}\label{eq:bareHamiltonian}
H_{0}=\sum_{\textbf{k},\sigma} \psi^{\dag}_{\textbf{k},\sigma}
H(\textbf{k})
\psi_{\textbf{k},\sigma},\nonumber\\
\end{align}
where
\begin{align}\label{eq:Hk}
H(\textbf{k})=
(\varepsilon_{+}(\textbf{k})-\mu)1+\varepsilon_{-}(\textbf{k})\tau_{3}+\varepsilon_{xy}(\textbf{k})\tau_{1}.
\nonumber\\
\end{align}
Here a two-component field $\psi^{\dag}_{\textbf{k},\sigma}$
= $[d^{\dag}_{xz,\sigma}(\textbf{k}),d^{\dag}_{yz,\sigma}(\textbf{k})]$
describes the creation of particles with $d_{xz}$ and $d_{yz}$ orbital flavors with spin $\sigma$,
and the Pauli matrix $\tau$ connects these two orbital states.
In the above,
\begin{align}\label{eq:Hterms}
\varepsilon_{+}(\textbf{k})&=-(t_{1}+t_{2})(\cos k_{x} + \cos k_{y}) - 4 t_{3}\cos k_{x}\cos k_{y},\nonumber\\
\varepsilon_{-}(\textbf{k})&=-(t_{1}-t_{2})(\cos k_{x} - \cos k_{y}),\nonumber\\
\varepsilon_{xy}(\textbf{k})&=- 4 t_{4}\sin k_{x}\sin k_{y}.\nonumber\\
\end{align}

Diagonalization of $H(\textbf{k})$ gives rise to the following two band dispersions,

\begin{align}\label{eq:banddispersion}
E_{\pm}(\textbf{k})=
\varepsilon_{+}(\textbf{k})-\mu \pm \sqrt{\varepsilon^{2}_{-}(\textbf{k})+\varepsilon^{2}_{xy}(\textbf{k})}.
\nonumber\\
\end{align}

In addition to time-reversal symmetry $T$,
the Hamiltonian $H_{0}$ has the $C_{4}$ point group symmetry, which
consists of the four-fold rotation $C_{\frac{\pi}{2}}$, the inversion $I$,
and the two
reflections $P_{x}$ and $P_{y}$ mapping $x$ to $-x$ and $y$ to $-y$, respectively.
Each symmetry operation
transforms a two-component field $\psi_{\sigma}(k_{x},k_{y})$
in the following way,

\begin{align}\label{eq:symmetrytransform}
\quad C_{\frac{\pi}{2}} :
\psi_{\sigma}(k_{x},k_{y}) &\rightarrow i\tau_{2}\psi_{\sigma}(-k_{y},k_{x}),\nonumber\\
\quad P_{x} :
\psi_{\sigma}(k_{x},k_{y}) &\rightarrow -\tau_{3}\psi_{\sigma}(-k_{x},k_{y}),\nonumber\\
\quad P_{y} :
\psi_{\sigma}(k_{x},k_{y}) &\rightarrow \tau_{3}\psi_{\sigma}(k_{x},-k_{y}),\nonumber\\
\quad I :
\psi_{\sigma}(k_{x},k_{y}) &\rightarrow -\tau_{0}\psi_{\sigma}(-k_{x},-k_{y}).\nonumber\\
\end{align}

If we choose the hopping parameters in such a way as
$t_{1}$=-1.0, $t_{2}$=1.3,  $t_{3}$ = $t_{4}$ = -0.85,
the $H_{0}$ works as an effective two-band Hamiltonian
describing the Fe-pnictide systems.~\cite{Raghu}
Given the hopping parameters above, the Fermi surface
consists of two hole pockets and two electron pockets
when the system is near half-filling.~\cite{Raghu,Ran}
A pair of electron and hole pockets are connected
by a nesting wave vector $\textbf{Q}=(\pi,0)$ (or $(0,\pi)$),
which drives various density wave instabilities.\cite{Chubukov}
Here we choose $\textbf{Q}=(\pi,0)$~\cite{footnote_Q} and perform a detailed study about
the band structures of density wave ground states considering all possible
density wave order parameters.

\begin{figure}[t]
\centering
\includegraphics[width=8.5 cm]{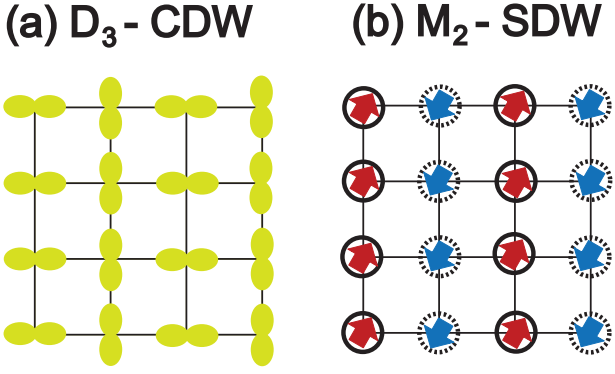}
\caption{(Color online)
Description of representative density wave ordering patterns
with the ordering wave vector $\textbf{Q}=(\pi,0)$.
(a) $D_{3}$ charge density wave ($D_{3}$-CDW) ordering.
$d_{xz}$ and $d_{yz}$ orbitals align alternatively
along the $x$ direction while local charge
and spin densities are uniform.
(b) $M_{2}$ spin density wave ($M_{2}$-SDW) ordering.
Here orbital and spin orderings occur at the same time.
A solid circle represents the $d_{+}$ orbital defined as
$d_{+}=(d_{xz}+id_{yz})/\sqrt{2}$ while
a dotted circle indicates the $d_{-}$ orbital given by
$d_{-}=(d_{xz}-id_{yz})/\sqrt{2}$.
The arrows inside circles describe spin ordering.
} \label{fig:orderparameters}
\end{figure}

\subsection{\label{sec:orderparameter} Symmetry of order parameters}
We consider various on-site density wave order parameters and investigate
their symmetry properties.
Since we have two orbitals per site, there are 4 different
on-site charge density wave (CDW) states with the ordering wave vector $\textbf{Q}=(\pi,0)$,
which are given by

\begin{align}
\hat{D}_{i}=\frac{1}{N}\sum_{\textbf{k}}\sum_{a,b=1}^{2}
d^{\dag}_{a,\sigma}(\textbf{k})[\tau_{i}]_{ab} d_{b,\sigma}(\textbf{k}+\textbf{Q}),
\end{align}
where $a$, $b$ are indices describing the $d_{xz}$ ($a$=1) or $d_{yz}$ ($a$=2) orbital states.
$N$ counts the number of unit cells in the system.
Similarly, we also define spin density wave (SDW) states choosing the spin ordering
direction along the $z$-axis,

\begin{align}
\hat{M}_{i}=\frac{1}{N}\sum_{\textbf{k}}\sum_{a,b=1}^{2}
d^{\dag}_{a,\sigma_{1}}(\textbf{k})[\tau_{i}]_{ab}[s^{z}]_{\sigma_{1} \sigma_{2}} d_{b,\sigma_{2}}(\textbf{k}+\textbf{Q}).
\end{align}

These 8 order parameters represent distinct phases
with different broken symmetries. For example,
the $D_{3}$ CDW order parameter corresponds to $\sum_{i} (-1)^{i_x}  (n_{i,xz} - n_{i, yz}) $ where $n_{i, a}$ is the density of electrons
with the orbital $a$ at the site $i$. Thus it is characterized by the relative density difference between
two orbitals (orbital ordering), which alternates along the $x$ direction,
while keeping the total density $(n_{i, xz}+n_{i, yz})$ uniform on every site as shown in Fig.~\ref{fig:orderparameters} (a).
This breaks translational symmetry doubling the unit cell along $x$ direction.
On the other hand, the $M_{2}$ SDW order parameter described in Fig.~\ref{fig:orderparameters} (b)
corresponds to a staggered spin-orbit coupling.
This is because $M_{2}$ can be written as
$\sum_{i}(-1)^{i_x} S_{i,z} L_{i,z}$ where $L_{i,z}$ is proportional to
$(d_{i,xz}+ i d_{i,yz})^{\dagger} (d_{i,xz}+i d_{i,yz}) - (d_{i,xz} - i d_{i,yz})^\dagger (d_{i,xz} - i d_{i,yz})
= 2 i d_{i,xz}^{\dagger} d_{i,yz} + h.c.$. However, unlike the uniform spin-orbital coupling $\sum_{i}S_{i,z} L_{i,z}$,
the $M_2$ is a staggered spin-orbit coupling with alternating signs along the $x$ direction.
It breaks spin-rotational and
translational symmetries but preserves time reversal symmetry.
In addition, $D_0$ and $M_{0}$ describe conventional charge and spin density wave states, respectively.
It was found that $M_{0}$ describes the leading density wave instability in Fe-pnictides.~\cite{Ran}

The above 8 order parameters can be distinguished by
their transformation properties under lattice symmetries.
The symmetries of density wave order parameters
are summarized in Table~\ref{table:order_symmetry}.
\begin{table}
\begin{tabular}{@{}|c|c|c|c|c|c|c|c|c|}
\hline \hline
& $D_{0}$ & $D_{1}$ & $D_{2}$ & $D_{3}$ & $M_{0}$  & $M_{1}$ & $M_{2}$ & $M_{3}$ \\
\hline \hline
$P_{x}$ & + & - & - & + & + & - & - & +  \\
$P_{y}$ & + & - & - & + & + & - & - & +  \\
$I$     & + & + & + & + & + & + & + & +  \\
$T$     & + & + & - & + & - & - & + & - \\
\hline \hline
\end{tabular}
\caption{Symmetry of density wave order parameters. Here `+' (`-')
indicates `even' (`odd') symmetry of the order parameters
under the corresponding symmetry operation.}
\label{table:order_symmetry}
\end{table}
Note that every density wave state has even parity under the inversion symmetry.
Moreover, all the diagonal density wave states $D_{i}$ (or $M_{i}$) with $i$ = 0 or 3
are even under the two reflection symmetries while the other off-diagonal
density wave states with $i$ = 1 or 2 are odd under the reflections.
These symmetry properties of density wave order parameters
strongly constrain the location of Dirac nodes generated by the Brillouin zone folding
and the winding numbers around Dirac nodes in the momentum space,
which are discussed in detail in the following section.

\section{\label{sec:NodalDW} Nodal density wave phases}

One intriguing property of the $\textbf{Q}=(\pi,0)$ density wave ground states
is that a large number of Dirac nodes emerge in the band structure.~\cite{Ran}
The numbers and locations
of the nodal points depend on band dispersions and the symmetries of the order parameters.

There are two different sources generating nodal points in general.
One way is via introducing a density wave order parameter carrying a finite momentum.
This induces a Brillouin zone folding which generates several band touching points.
In most cases, the degeneracy at the band touching point is lifted because
the density wave order parameter induces a finite overlap between the pair of states touching at a point.
Henceforth a band gap opens up. However, when the band touching occurs at
a high symmetry point in the Brillouin zone, the overlap matrix vanishes due to the lattice symmetries,
generating symmetry protected nodal points.

The second group of nodal points come from the splitting of quadratic band touching points,
which exist in the bare bandstructure.
Because of the underlying four-fold rotational symmetry, the original hopping Hamiltonian in Eq.(\ref{eq:bareHamiltonian})
supports quadratic band crossing points.~\cite{Ran, kaisun1, kaisun2, wen1, Vafek}
The introduction of the density wave order parameter carrying a finite momentum
splits a quadratic band touching point into two Dirac points along
high symmetry directions in the momentum space.
In the following, we discuss in detail the relation
between the order parameter symmetry and the locations of Dirac points
derived from these two different sources in separate subsections.

\subsection{\label{sec:zonefolding} Dirac nodes generated by Brillouin zone folding}

\begin{figure}[t]
\centering
\includegraphics[width=8.5 cm]{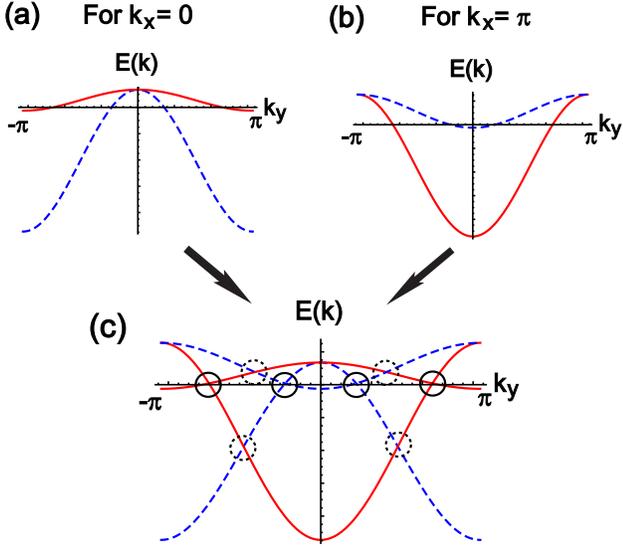}
\caption{(Color online)
(a) Band dispersion along $k_{x}=0$ for the hopping Hamiltonian in Eq.(\ref{eq:bareHamiltonian}).
Here we use a red solid (blue dotted) line to indicate a band which is odd (even)
under the $P_{x}$ reflection symmetry.
(b) Band dispersion along the $k_{x}$=$\pi$ direction.
(c) Band structure along $k_{x}$=0 after the unit cell doubling due to
the $\textbf{Q}$=($\pi$,0) density wave ordering. Four bands in (a) and (b) meet and disperse together
along $k_{x}$=0 after the Brillouin zone folding. Note that the zone folding
generates 8 band touching points. Black solid (dotted) circles indicate
the band touching points between two bands
having the same (opposite) $P_{x}$ eigenvalues.
} \label{fig:band_kx0}
\end{figure}

We first focus on the generation of Dirac nodes along the $k_{y}$-axis.
In Fig.~\ref{fig:band_kx0}(a) (Fig.~\ref{fig:band_kx0}(b)), we plot the energy dispersion of
the two bands given in Eq.~(\ref{eq:banddispersion}) along the $k_{x}=0$ ($k_{x}=\pi$) direction.
Since the $\varepsilon_{xy}$ term in Eq.~(\ref{eq:Hterms}),
which describes the hybridization between $d_{xz}$
and $d_{yz}$ orbitals, vanishes along the $k_{x}=0$ axis,
the upper and lower bands in Fig.~\ref{fig:band_kx0}(a)
are just $d_{yz}$and $d_{xz}$ bands, respectively.
For $k_{x}=0$, $H(\textbf{k})$ is invariant under the $P_{x}$ reflection symmetry
which transforms a momentum $k_{x}$ to $-k_{x}$.
Therefore each band
is an eigenstate of $P_{x}$ with eigenvalues of $\pm1$. This is consistent with
the fact that $d_{xz}$ ($d_{yz}$) orbital is odd (even) under $P_{x}$.
Similar analysis can also be applied to the two bands dispersing along the $k_{x}=\pi$ direction.
Since $H(\textbf{k})$ has $P_{x}$ symmetry along $k_{x}=\pi$,
the two bands also have definite $P_{x}$ eigenvalues.
In Fig.~\ref{fig:band_kx0}, the $P_{x}$ even (odd) bands are represented by
blue dotted (red solid) lines.

Once we introduce a density wave order parameter with the ordering wave vector $\textbf{Q}=(\pi,0)$,
the unitcell doubles along the $x$-direction, which leads to the Brillouin zone folding
in the momentum space.
Thus within the reduced Brillouin zone,
we have four bands dispersing along the $k_{y}$-axis.
Note that the zone folding generates 8 band touching points,
which are indicated by circles in Fig.~\ref{fig:band_kx0}(c).
Here the band touching point between two bands with the same (opposite) $P_{x}$
eigenvalues is encircled by a solid (dotted) circle.
\begin{figure}[t]
\centering
\includegraphics[width=6.5 cm]{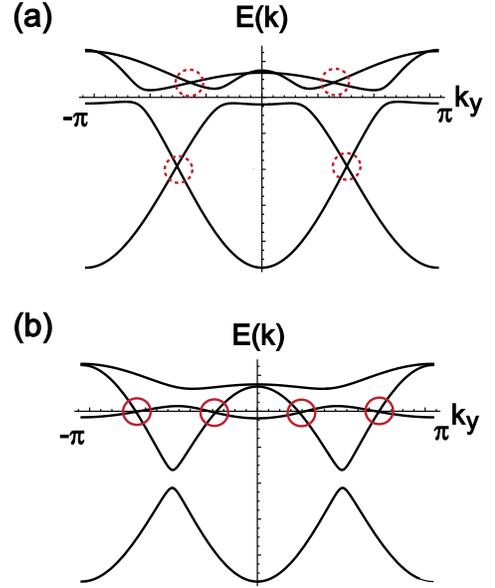}
\caption{(Color online)
The band structures of the $\textbf{Q}=(\pi,0)$ density wave ground states along the $k_{y}$-axis.
(a) For $P_{x}$ even density wave states.
(b) For $P_{x}$ odd density wave states.
} \label{fig:GappedBands_kx0}
\end{figure}

The degeneracy between two states,
$|\psi_{1}(\textbf{k})\rangle$ and $|\psi_{2}(\textbf{k}+\textbf{Q})\rangle$,
touching at the momentum $\textbf{k}$ after the Brillouin zone folding,
is lifted when the matrix element of the density wave order parameter $\hat{D}_{i}$
between these two states is finite,
that is, $\langle\psi_{1}(\textbf{k})|\hat{D}_{i}|\psi_{2}(\textbf{k}+\textbf{Q})\rangle\neq0$.
Therefore if the order parameter $\hat{D}_{i}$ (or $\hat{M}_{i}$) is $P_{x}$ even,
the degeneracy is lifted when the two degenerate bands have the same $P_{x}$ eigenvalues.
However, the nodal point remains gapless
if the two degenerate bands have the opposite $P_{x}$ eigenvalues.

On the other hand, if the density wave order parameter $\hat{D}_{i}$ (or $\hat{M}_{i}$)
is $P_{x}$ odd, the full Hamiltonian is not invariant under $P_{x}$ anymore.
However, even in this case
$\langle\psi_{1,s_{1}}(\textbf{k})|\hat{D}_{i}|\psi_{2,s_{2}}(\textbf{k}+\textbf{Q})\rangle=0$
in the weak coupling limit,
if $s_{1}=s_{2}$ where $s$ refers to $P_{x}$ eigenvalues.
Namely, the matrix element of $D_{i}$, which is odd under $P_{x}$,
vanishes when the two degenerate eigenstates have the same $P_{x}$ eigenvalues.
To understand this point clearly, let us define the eigenvector $|\phi^{(0)}_{n,s}(\textbf{k})\rangle$
of the hopping Hamiltonian $H_{0}$ with the even ($s$ = +) or odd ($s$ = -) $P_{x}$ eigenvalues.
Here $n$ is a band index.
Now we turn on a small density wave order parameter $\hat{D}_{i}$
which is odd under $P_{x}$. Since $P_{x}$ eigenvalue is not a good quantum number,
$|\phi^{(0)}_{n,s}(\textbf{k})\rangle$ can be contaminated
by the states with the opposite $P_{x}$ eigenvalue
$|\phi^{(0)}_{m,\bar{s}}(\textbf{k+Q})\rangle$, leading to

\begin{displaymath}
|\phi^{(0)}_{n,s}(\textbf{k})\rangle \mathop{\longrightarrow}^{\hat{D}_{i}}
|\psi_{n,s}(\textbf{k})\rangle=|\phi_{s}(\textbf{k})\rangle+|\phi_{\bar{s}}(\textbf{k+Q})\rangle,
\end{displaymath}
where $|\phi_{s}(\textbf{k})\rangle=\sum_{n}c_{n}|\phi^{(0)}_{n,s}(\textbf{k})\rangle$ is a linear combination of the states
with the $P_{x}$ eigenvalue of $s$,
while $|\phi_{\bar{s}}(\textbf{k+Q})\rangle=\sum_{n}c'_{n}|\phi^{(0)}_{n,\bar{s}}(\textbf{k+Q})\rangle$ is a linear combination
of the states
with the opposite $P_{x}$ eigenvalue of $\bar{s}$.
Notice that $|\phi_{s}(\textbf{k})\rangle$ and $|\phi_{\bar{s}}(\textbf{k+Q})\rangle$
have a momentum difference given by the ordering wave vector $\textbf{Q}$ carried by the density wave
order parameter $\hat{D}_{i}$.
Because of the fact that the two components of the wave function with the opposite $P_{x}$ eigenvalues have
the momentum difference given by $\textbf{Q}$,
it is straight forward to show that
$\langle\psi_{1,s_{1}}(\textbf{k})|\hat{D}_{i}|\psi_{2,s_{2}}(\textbf{k}+\textbf{Q})\rangle=0$
if $s_{1}=s_{2}$.

Therefore the nodal point remains gapless if the order parameter $\hat{D}_{i}$
is $P_{x}$ even ($P_{x}$ odd) while the two generated bands have the opposite (same)
$P_{x}$ eigenvalues. It means that 4 nodal
points among the 8 band touching points remain gapless
independent of the condition that the order parameter is even or odd under $P_{x}$ reflection symmetry.

In Fig.~\ref{fig:GappedBands_kx0} we plot the band structure
of the density wave ground states along the $k_{y}$-axis.
Fig.~\ref{fig:GappedBands_kx0} (a) corresponds to
the density wave orders $D_{0}$, $D_{3}$, $M_{0}$, $M_{3}$,
which are $P_{x}$ even,
while Fig.~\ref{fig:GappedBands_kx0} (b) describes
the band structure for the other order parameters,
$D_{1}$, $D_{2}$, $M_{1}$, $M_{2}$,
which are odd under $P_{x}$ symmetry.
Notice that nodal points show opposite behavior
for these two different classes of order parameters.
Namely, when a nodal points remains gapless for one order parameter,
it is gapped out for the other order parameter which has the opposite $P_{x}$ eigenvalue.

\begin{figure}[t]
\centering
\includegraphics[width=8.5 cm]{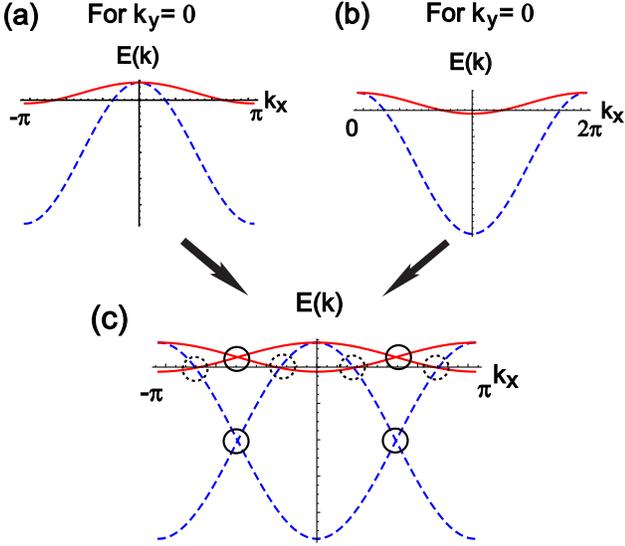}
\caption{(Color online)
(a) Band dispersion along $k_{y}=0$ centered at $\textbf{k}$=(0,0).
(b) Band dispersion along $k_{y}=0$ centered at $\textbf{k}$=($\pi$,0).
(c) Band structure along $k_{y}$=0 after the unit cell doubling.
Four bands in (a) and (b) meet and disperse together
along $k_{y}$=0 after the Brillouin zone folding.
Here we use a red solid (blue dotted) line to indicate a band which is odd (even)
under the $P_{y}$ symmetry.
Black solid (dotted) circles indicate
the band touching points between two bands
having the same (opposite) $P_{y}$ eigenvalues.
} \label{fig:band_ky0}
\end{figure}

We can extend the same analysis to understand nodal points lying along the $k_{x}$-axis.
In this case we have $P_{y}$ reflection symmetry mapping $y$ to $-y$.
However, compared to the previous analysis for nodal points on the $k_{y}$-axis,
there is one important difference in this case.
Before the unitcell doubling, we have two bands dispersing along the $k_{x}$-axis.
The Brillouin zone folding induces overlaps of these two bands with themselves.
In Fig.~\ref{fig:band_ky0}, we plot the dispersion of the two bands along the $k_{x}$-axis
centered at $\textbf{k}=(0,0)$ (Fig.~\ref{fig:band_ky0}(a)) and
at $\textbf{k}=(\pi,0)$ (Fig.~\ref{fig:band_ky0}(b)).
The 4 bands after the zone folding displayed in Fig.~\ref{fig:band_ky0}(c)
can be obtained by superposing
the 4 bands in Fig.~\ref{fig:band_ky0}(a) and (b).
In Fig.~\ref{fig:band_ky0}(c), we plot the bandstructure
from $k_{x}=-\pi$ to $k_{x}=\pi$ for convenience although
the first Brillouin zone is from $k_{x}=-\pi/2$ to $k_{x}=\pi/2$.
Note that in Fig.~\ref{fig:band_ky0}(c) the location of solid and dotted circles
are interchanged compared to those in Fig.~\ref{fig:band_kx0}(c).
Because of this difference, the location of Dirac nodes along the $k_{x}$
and $k_{y}$ axes also show the opposite behaviors.

\subsection{\label{sec:quadratic} Dirac nodes from quadratic band crossing}

The band structure of the two-band hopping Hamiltonian
$H_{0}$ in Eq.~(\ref{eq:bareHamiltonian}) supports two quadratic band crossing
points at $\textbf{k}=(0,0)$ and $\textbf{k}=(\pi,\pi)$.~\cite{Ran, Qi1}
Splitting of these quadratic band crossing points generates additional Dirac points,
which contribute additional Chern numbers for various insulating phases.

Expanding the Hamiltonian $H(\textbf{k})$ in Eq.~($\ref{eq:Hk}$)
near $\textbf{k}=(0,0)$, we obtain the following
low energy effective Hamiltonian,

\begin{align}\label{eq:Hquad}
\text{H}_{\text{eff}}=\int d^{2}k \psi^{\dag}(\textbf{k})
H_{\text{quad}}(\textbf{k})
\psi(\textbf{k}),
\end{align}
in which
\begin{align}\label{eq:Hquad2}
H_{\text{quad}}(\textbf{k})=
\alpha(k^{2}_{x}+k^{2}_{y})\hat{\tau}_{0}+\beta k_{x}k_{y}\hat{\tau}_{1}+\gamma (k^{2}_{x}-k^{2}_{y})\hat{\tau}_{3},
\end{align}
where $\alpha=(t_{1} + t_{2} + 4 t_{3})/2$, $\beta=-4 t_{4}$,
and $\gamma=(t_{1} - t_{2})/2$.
Nontrivial topological property of the quadratic band crossing point
is reflected in the winding number $N_{\text{w}}$, which is defined as,~\cite{kaisun2}
\begin{align}\label{eq:winding}
N_{\text{w}}\equiv \frac{1}{\pi i}\oint_{\textbf{C}} d\textbf{k}\cdot \langle\psi^{\dag}(\textbf{k})|\nabla_{\textbf{k}}
|\psi(\textbf{k})\rangle,
\end{align}
where $|\psi(\textbf{k})\rangle$ is a Bloch wave function corresponding
to one of the bands involved in the band touching and $\textbf{C}$
is a closed loop in the momentum space encircling the band crossing point.
A quadratic band crossing point contributes $N_{W} = \pm 2$,
which is twice larger than the winding number around a Dirac point.~\cite{MOnoda,kaisun1,kaisun2,wen1}

Adding a generic perturbation given by
$V=\sum^{3}_{i=1}m_{i}\hat{\tau}_{i}$, the degeneracy at the quadratic
band crossing point can be lifted.
$m_{2}$ term breaks time-reversal symmetry and the degeneracy is lifted
by opening a gap. On the other hand, $m_{1}$ and $m_{3}$ terms that break
4-fold rotational symmetry, split the quadratic band touching point
into two Dirac points.~\cite{kaisun1,kaisun2,wen1}

Now we consider the effect of the $\textbf{Q}=(\pi,0)$ density wave orderings
on the degeneracy lifting at quadratic band crossing points.
Since the density wave order parameters carry the momentum $\textbf{Q}=(\pi,0)$,
they cannot couple to the degenerate states at $\textbf{k}=(0,0)$
(or $\textbf{k}=(\pi,\pi)$) at first order.
The lowest order contribution to degeneracy lifting at quadratic band touching points
starts from the second order processes.
We first consider charge density wave order parameters given by,

\begin{align}\label{eq:CDW}
\text{H}_{\text{CDW}}=\sum_{\textbf{k},\sigma} \psi^{\dag}_{\textbf{k},\sigma}\hat{D}
\psi_{\textbf{k
}+\textbf{Q},\sigma},
\end{align}
where $\hat{D}=\sum^{3}_{i=0}D_{i}\hat{\tau}_{i}$.
Treating the above H$_{\text{CDW}}$ as a perturbation,
the standard second order perturbation theory gives rise to the following
effective Hamiltonian near the quadratic band touching point at $\textbf{k}=(0,0)$,

\begin{align}
H^{\text{eff}}_{\text{quad}}=&H_{\text{quad}}(\textbf{k}) + H_{\text{mass}},
\nonumber\\
=&H_{\text{quad}}(\textbf{k}) + \sum^{3}_{i=0}m^{(i)}_{\Gamma}\hat{\tau}_{i},
\end{align}
in which
\begin{align}\label{eq:massquad}
m^{(0)}_{\Gamma}=&\lambda
\{(D^{2}_{1}+D^{2}_{2})(t_{1}+t_{2}+4t_{3})\nonumber\\
&+(D_{0}+D_{3})^{2}(t_{2}+2t_{3})+(D_{0}-D_{3})^{2}(t_{1}+2t_{3})\},
\nonumber\\
m^{(1)}_{\Gamma}=&2\lambda
D_{1}\{D_{3}(-t_{1}+t_{2})+D_{0}(t_{1}+t_{2}+4t_{3})\},
\nonumber\\
m^{(2)}_{\Gamma}=&2\lambda
D_{2}\{D_{3}(-t_{1}+t_{2})+D_{0}(t_{1}+t_{2}+4t_{3})\},
\nonumber\\
m^{(3)}_{\Gamma}=&\lambda
\{(D^{2}_{1}+D^{2}_{2})(t_{1}-t_{2})\nonumber\\
&+(D_{0}+D_{3})^{2}(t_{2}+2t_{3})-(D_{0}-D_{3})^{2}(t_{1}+2t_{3})\},
\nonumber\\
\end{align}
where $\lambda=-1/\{8(t_{1}+2t_{3})(t_{2}+2t_{3})\}$.
Note that as long as only one of the order parameters has finite magnitude
while all the other order parameters are zero,
$m^{(1)}_{\Gamma}=m^{(2)}_{\Gamma}=0$.
In other words, if $D_{n}\neq 0$ for a given $n$ while all the other $D_{i\neq n}=0$,
the quadratic band crossing point always splits into two Dirac points along the main axes.

\begin{figure}[t]
\centering
\includegraphics[width=8.5 cm]{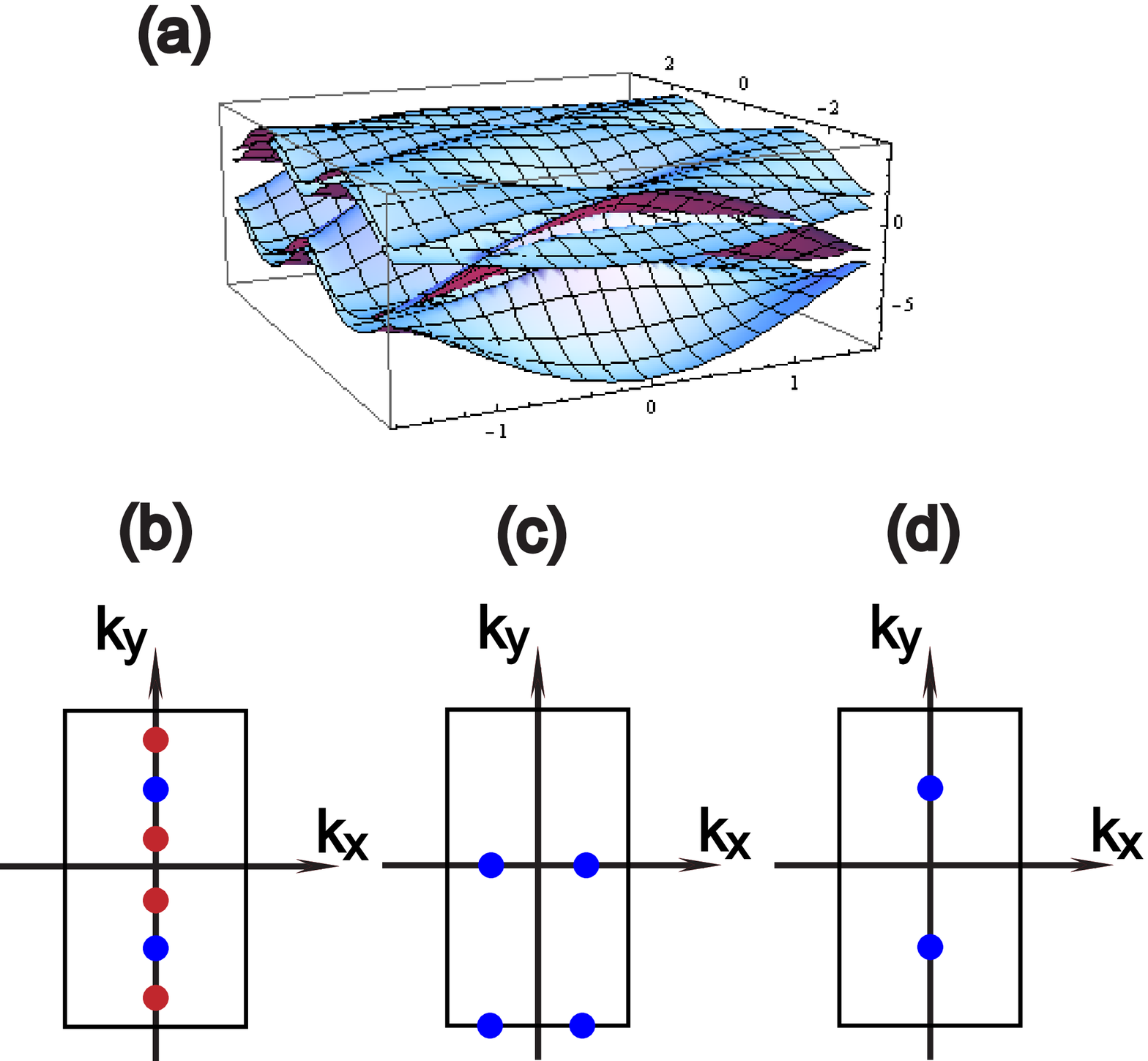}
\caption{(Color online)
Distribution of Dirac points for the $D_{3}$ (or $M_{3}$) density wave state.
(a) Four bands within the reduced Brillouin zone.
The energy eigenvalue $E_{i}(\textbf{k})$ for a band $i$ satisfies
$E_{1}(\textbf{k})\geq E_{2}(\textbf{k})\geq E_{3}(\textbf{k}) \geq E_{4}(\textbf{k})$.
(b) Dirac points between the band 1 and 2,
(c) between band 2 and 3,
(d) between band 3 and 4.
Blue (Red) dots indicates Dirac points coming from the Brillouin zone folding
(splitting quadratic band touching points).
} \label{fig:nodelocations}
\end{figure}

Combining the contributions both from the Brillouin zone folding
and from the splitting of quadratic band crossing points,
we show the distribution
of Dirac points for a $D_{3}$ (or $M_{3}$) density wave ground state
in Fig.~\ref{fig:nodelocations}.
There are four bands within the reduced Brillouin zone as shown in Fig.~\ref{fig:nodelocations}(a).
We assign a band index $i$ such that the energy eigenvalue $E_{i}(\textbf{k})$
of the band $i$ satisfies
$E_{1}(\textbf{k})\geq E_{2}(\textbf{k})\geq E_{3}(\textbf{k}) \geq E_{4}(\textbf{k})$.
The location of Dirac points between the upper two bands (band $1$ and $2$)
are indicated in Fig.~\ref{fig:nodelocations}(b).
Similarly, the Dirac points between the middle (bottom) two bands
are described in Fig.~\ref{fig:nodelocations}(c) (Fig.~\ref{fig:nodelocations}(d)).
Notice that there are many Dirac touching points between the bands.
Blue dots indicate the nodal points coming from the Brillouin zone folding
induced by the $\textbf{Q}=(\pi,0)$ density wave states.
On the other hand, red dots result from the splitting of quadratic band touching points.
Two quadratic band touching points generate four Dirac points lying on the
$y$-axis.
With the understanding of the origin and locations of Dirac points,
below we discuss how to achieve TDW insulators.

\section{\label{sec:TopologicalPhase} Topological properties of the gapped density wave phases}

A single density wave order parameter induces a metallic phase
with many Dirac points.
The locations of Dirac points are determined by the transformation properties
of the order parameters under the reflection symmetries $P_{x}$ and $P_{y}$.
Therefore to get an insulating phase,
two coexisting density wave states,
in which one is even and the other is odd under the $P_{x}$ and $P_{y}$ symmetries,
are required.
In addition, according to the order parameter symmetries summarized in Table~\ref{table:order_symmetry},
if time-reversal invariance is imposed,
there are only four different ways of choosing a pair of density wave order parameters,
which give rise to a gapped phase.
The four pairs of time reversal invariant density wave order parameters with the opposite
transformation properties under the reflections $P_{x}$ and $P_{y}$, are given by
($D_{3}$, $D_{1}$), ($D_{3}$, $M_{2}$), ($D_{0}$, $D_{1}$), and ($D_{0}$, $M_{2}$).

Since the $z$ component of the spin, $S_{z}$ is conserved,
the Chern number $N^{(n)}_{C,\sigma}$
is well defined for each band in a fully gapped phase.~\cite{SpinCh1,SpinCh2,Fu-Kane2}
Here $N^{(n)}_{C,\uparrow}$ ($N^{(n)}_{C,\downarrow}$) is
the Chern number of the $n$th spin-up (spin-down) band.
For every pair of the density wave order parameters generating a fully gapped phase,
the four bands within the reduced Brillouin zone are well-separated
from each other with a finite gap between any pairs of the bands.
Each band is distinguished by the index $n$ ranging from 1 to 4 as the energy decreases.
The Chern number $N^{(n)}_{C,\sigma}$ of the $n$th band with the spin $\sigma$
is defined as,
\begin{align}\label{eq:chernintegral}
N^{(n)}_{C,\sigma}=&\frac{1}{2\pi}\int_{\text{RBZ}} d^{2}k F^{(n)}_{\sigma}(\textbf{k}),
\end{align}
where the momentum space Berry curvature $F^{(n)}_{\sigma}(\textbf{k})$
for the $n$th band with the spin $\sigma$ is defined as
$F^{(n)}_{\sigma}(\textbf{k})\equiv\partial_{k_{x}}A^{(n)}_{y,\sigma}(\textbf{k})-\partial_{k_{y}}A^{(n)}_{x,\sigma}(\textbf{k})$
in which the Berry potential $A^{(n)}_{\mu,\sigma}(\textbf{k})$ is given by
$A^{(n)}_{\mu,\sigma}(\textbf{k})=
-i\langle\Phi^{(n)}_{\sigma}(\textbf{k})|\partial_{k_{\mu}}|\Phi^{(n)}_{\sigma}(\textbf{k})\rangle$.~\cite{TKNN1,TKNN2}
Here the Bloch wave function $|\Phi^{(n)}_{\sigma}(\textbf{k})\rangle$ is
defined within the reduced Brilluin zone (RBZ).

\begin{table}
\begin{tabular}{@{}|c|c|c|}
\hline \hline
& $N_{C,\uparrow}$ & $N_{C,\downarrow}$\\
\hline \hline
band 1 & + 1 & - 1  \\
band 2 & - 3 & + 3  \\
band 3 & + 3 & - 3  \\
band 4 & - 1 & + 1 \\
\hline \hline
\end{tabular}
\caption{The Chern number of each band for
the gapped density wave ground state
with finite $D_{3}$ and $M_{2}$.}
\label{table:Chernnumbers}
\end{table}

Explicit computation of the Chern numbers using Eq.~(\ref{eq:chernintegral}) shows that
every band of the insulating density wave phase with finite $D_{3}$
and $M_{2}$, has a nonzero Chern number as shown in Table~\ref{table:Chernnumbers}.
On the other hand,
every band has zero Chern number for the other
three gapped phases defined with a pair of nonzero order parameters
given by
($D_{3}$, $D_{1}$), ($D_{0}$, $D_{1}$), and ($D_{0}$, $M_{2}$).

\subsection{\label{sec:Chernnumber} Topological properties of the topological density wave ground state
with finite $D_{3}$ and $M_{2}$}
\begin{figure}[t]
\centering
\includegraphics[width=8.5 cm]{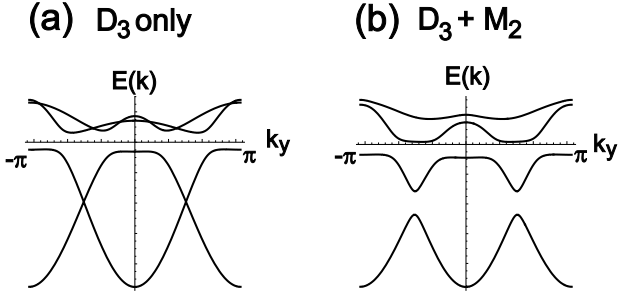}
\caption{(Color online)
(a) Band structure of the $D_{3}$ nodal density wave state
along the $k_{y}$ axis.
(b) Band structure of the gapped density wave phase
with finite $D_{3}$ and $M_{2}$.
} \label{fig:D3M2_kx0}
\end{figure}

Nontrivial topological properties of the fully gapped density wave phase with nonzero $D_{3}$ and $M_{2}$
can be understood in the following way.
We first consider the charge density wave state with the finite
$D_{3}$ order parameter.
The $D_{3}$ density wave phase supports many Dirac points
whose distribution is described in Fig.~\ref{fig:nodelocations}.
Now we turn on a small $M_{2}$ which induces gap opening at each nodal point,
leading to the fully gapped insulating phase, which is described in Fig.~\ref{fig:D3M2_kx0}.
The degeneracy lifting at each nodal point can be understood
as a result of the mass perturbation, induced by the finite $M_{2}$,
to the gapless Dirac particles.
The nonzero Chern number of each band is obtained by adding up
the Chern number contributions of the massive Dirac particles
derived from the corresponding band.

We first focus on the two Dirac nodes lying along the $k_{y}$ axis
between the band 3 and 4 shown in Fig.~\ref{fig:nodelocations}(d).
The effective Dirac Hamiltonian can be obtained by linearizing
the Hamiltonian near the two nodal points sitting at the momentum $\textbf{k}$=$\textbf{k}_{+}$ and $\textbf{k}_{-}$. To simplify
the computational procedures, the hopping parameters $t_{i}$ are slightly shifted
from the initial values given in Sec.~\ref{sec:tightbinding}
to $t_{1}=-t_{2}=t_{3}=t_{4}$=-1.0.
This small parameter change does not affect the topological properties
of the gapped phase but shifts the nodal points to $\textbf{k}_{\pm}=(0,\pm \frac{\pi}{2})$
making analytical analysis simpler.
The effective Hamiltonian describing the low energy fermions near these two nodal points
is given by
\begin{align}\label{eq:HDirac}
\text{H}^{\text{Dirac}}_{\text{eff}}=\sum_{\mu=\pm}\int d^{2}q \Psi^{\dag}_{\mu,\sigma}(\textbf{q})
H^{\text{Dirac}}_{\mu,\sigma}(\textbf{q})
\Psi_{\mu,\sigma}(\textbf{q}),
\end{align}
where
\begin{align}\label{eq:HDirac2}
H^{\text{Dirac}}_{\mu,\sigma}(\textbf{q})=
-\mu\{(2+\frac{8}{\sqrt{4+D^{2}_{3}}}) q_{y}\hat{\tau}_{3}+\frac{4|D_{3}|}{\sqrt{4+D^{2}_{3}}} q_{x}\hat{\tau}_{1}\}.
\end{align}
Here the momentum $\textbf{q}$ of the Hamiltonian $H^{\text{Dirac}}_{\mu,\sigma}(\textbf{q})$
is measured with respect to the degeneracy point at $\textbf{k}=\textbf{k}_{\mu}$ ($\mu=+, -$).
The two-component fermion field $\Psi_{\mu,\sigma}(\textbf{q})$ is given by

\begin{equation}\label{eq:Diracfield}
\Psi_{\mu,\sigma}(\textbf{q}) =
\left( \begin{array}{cc}
\alpha_{1}d_{yz,\sigma}(\textbf{k}_{\mu}+\textbf{q})+\alpha_{2}d_{yz,\sigma}(\textbf{k}_{\mu}+\textbf{Q}+\textbf{q})  \\
\beta_{1}d_{xz,\sigma}(\textbf{k}_{\mu}+\textbf{q})+\beta_{2}d_{xz,\sigma}(\textbf{k}_{\mu}+\textbf{Q}+\textbf{q})
\end{array} \right)
\end{equation}
where the constant coefficients $\alpha_{i}$ and $\beta_{i}$ satisfy
$\alpha^{2}_{1}+\alpha^{2}_{2} = \beta^{2}_{1}+\beta^{2}_{2} = 1$.
Explicitly, $\alpha_{i}$ and $\beta_{i}$ ($i$=1,2) are given by
\begin{align}
\alpha_{1}&=(2+\sqrt{4+D^{2}_{3}})/\sqrt{8+2D^{2}_{3}+4\sqrt{4+D^{2}_{3}}},
\nonumber\\
\alpha_{2}&=D_{3}/\sqrt{8+2D^{2}_{3}+4\sqrt{4+D^{2}_{3}}},
\nonumber\\
\beta_{1}&=(2-\sqrt{4+D^{2}_{3}})/\sqrt{8+2D^{2}_{3}-4\sqrt{4+D^{2}_{3}}},
\nonumber\\
\beta_{2}&=D_{3}/\sqrt{8+2D^{2}_{3}-4\sqrt{4+D^{2}_{3}}}.
\nonumber
\end{align}
Notice that the first (second) component of $\Psi_{\mu,\sigma}(\textbf{q})$
is derived entirely from the $d_{yz}$ ($d_{xz}$) orbital.

Now we include the $M_{2}$ spin density wave order parameter which
generates a mass term in the low energy limit given by
\begin{align}\label{eq:Hmass1}
\text{H}^{\text{mass}}_{M_{2}}=\sum_{\mu=\pm}\int d^{2}q \Psi^{\dag}_{\mu,\sigma}(\textbf{q})
\{-\frac{2|D_{3}|M_{2}\sigma}{D_{3}\sqrt{4+D^{2}_{3}}} \hat{\tau}_{2}\}
\Psi_{\mu,\sigma}(\textbf{q}).
\end{align}
For the given spin $\sigma$, the mass term has the same magnitude and sign
at the two Dirac points.
At each Dirac point, this mass term opens a gap
and contributes to the Chern number $N_{C,\sigma}=+\frac{M_{2}\sigma}{2|M_{2}|}$
for the upper band (band 3) and
$N_{C,\sigma}=-\frac{M_{2}\sigma}{2|M_{2}|}$
for the lower band (band 4).~\cite{MOnoda, TBI_review, bernevig1, Redlich}
Adding the Chern number contributions from the two Dirac points,
the total Chern number of the band 4 with the spin $\sigma$
is given by $N^{(4)}_{C,\sigma}=-\text{sgn}(M_{2})\sigma$,
which is consistent with the result obtained from the integration of the
Berry curvature over the reduced Brillouin zone using Eq.~(\ref{eq:chernintegral}).
(See Table~\ref{table:Chernnumbers}.)
In the case of the band 3, the Chern number is determined
after including the additional contributions from the nodal points between
the band 2 and band 3.

Similarly, the trivial topological property of the gapped phase with finite $D_{3}$ and $D_{1}$
can also be understood by applying the same analysis.
For the $D_{3}$ nodal density wave state,
the small $D_{1}$ charge density wave order parameter
generates mass perturbations to Dirac particles, which can be described by the following Hamiltonian,
\begin{align}\label{eq:Hmass2}
\text{H}^{\text{mass}}_{D_{1}}=\sum_{\mu=\pm}\int d^{2}q \Psi^{\dag}_{\mu,\sigma}(\textbf{q})
\{\frac{2|D_{3}|D_{1}}{D_{3}\sqrt{4+D^{2}_{3}}} \hat{\tau}_{1}\}
\Psi_{\mu,\sigma}(\textbf{q}).
\end{align}
Notice that this term just induces the shifting of the nodal points away from the $k_{y}$-axis.
Once a Dirac point moves away from the reflection symmetry axis,
the degeneracy at the band touching point is no longer protected by the symmetry and a gap opens,
because the density wave order parameters support finite
matrix elements between the two bands touching at the nodal point.
Since the $D_{1}$ charge density wave order parameter
does not generate a mass term to the Dirac Hamiltonian,
it has no contribution to the Chern number leading
to the zero Chern numbers of all bands.

We apply similar analysis to every Dirac point
derived from the Brillouin zone folding for
all pairs of density wave order parameters
generating fully gapped phases.
In all cases, it is confirmed that the Chern number of each band obtained
by summing up the Chern number contributions from the Dirac points
is identical to the result obtained by
the integration of the Berry curvature in the momentum space.

In addition, the coexisting $D_{3}$ and $M_{2}$ density wave order parameters
also lift the degeneracies of the two quadratic band touching points
at $\textbf{k}=(0,0)$ and $(0,\pi)$ leading to a fully gapped bandstructure.
In contrast to the case of Dirac points, the Chern number obtained
by lifting a quadratic band degeneracy is
two times larger than the contribution from a single Dirac point.
However, since the two quadratic band crossing points
lead to the Chern number contributions with the opposite signs,
the net effect of the two quadratic band touching points vanishes.

\begin{table}
\begin{tabular}{@{}|c|c|c|}
\hline \hline
Band filling & Spin Chern Number ($C_{S}$)  & $Z_{2}$ invariant ($\nu$)\\
\hline \hline
$\frac{3}{4}$ & - 2 & 1  \\
$\frac{1}{2}$ & + 4 & 0  \\
$\frac{1}{4}$ & - 2 & 1  \\
\hline \hline
\end{tabular}
\caption{Spin Chern numbers and topological $Z_{2}$ invariants
for the gapped density wave ground state with finite $D_{3}$ and $M_{2}$.}
\label{table:z2invariant}
\end{table}
Since the four bands are well separated from each other
for the topological density wave phase with $D_{3}\neq 0$ and $M_{2}\neq 0$,
if the magnitude of the order parameter $M_{2}$ is large enough,
an insulating ground state is obtained whenever the Fermi level
lies in the gap between two neighboring bands.
Therefore there are three different insulating phases, in principle,
whenever the Fermi level lies between the band $n$ and $n+1$ ($n$=1, 2, 3) corresponding
to the band filling factor $N_{\text{filling}}=(4-n)/4$.
The topological property of the insulating phase
can be explicitly characterized by computing topological invariants.
We first consider the spin Chern number $C_{S}$ which is defined in the following way,
\begin{align}
C_{S}=\sum_{n\in occ}\{N^{(n)}_{C,\uparrow}-N^{(n)}_{C,\downarrow}\},
\end{align}
where the summation includes all the occupied bands.
When the $S_{z}$ is conserved, the spin Chern number $C_{S}$
is quantized and characterizes the two dimensional topological insulators.~\cite{SpinCh1}
In Table.~\ref{table:z2invariant} we show the spin Chern numbers for
the insulating phases. It is interesting that the spin Chern numbers
are nonzero for all cases. Therefore as long as the $z$ component of
the spin is conserved, we can obtain the topological insulator
with finite spin hall conductivity for every quarter filling.
However, in the presence of
spin non-conserving perturbations and disorders,
the spin Chern number is not well-defined and conserved only modulo 4.~\cite{Fu-Kane2,Fukui}
In other words, when the $S_{z}$ is not conserved, the half-filled system is equivalent to
the phase with zero spin Chern number, which is nothing but
a topologically trivial band insulator.
However, it is important to notice that even in this case the system remains as a topological insulator with
nonzero $Z_{2}$ topological invariant for 1/4 and 3/4 fillings.

We also compute the $Z_{2}$ topological invariant $\nu$ shown in Table~\ref{table:z2invariant},
which can be used to distinguish topological insulators ($\nu=1$)
from trivial band insulators ($\nu=0$)
for generic time-reversal invariant systems.
Since the system has the inversion symmetry, the $Z_{2}$ invariant
can be obtained from the parity eigenvalues $\xi_{m}(\Gamma_{l})$ of the occupied bands
with the band indices $m$
at the time-reversal invariant momenta $\bf{\Gamma}_{l}$.~\cite{Fu-Kane}
Using the reciprocal lattice vectors $\textbf{G}_{i}$ ($i$=1, 2),
the four time reversal invariant momenta can be written as $\bf{\Gamma}_{l=n_{1}n_{2}}$
=$(n_{1}\textbf{G}_{1}+n_{2}\textbf{G}_{2})/2$ with $n_{1,2}= 0, 1$.
Explicitly the $Z_{2}$ topological invariant $\nu$ is given by
\begin{align}\label{eq:strongnu}
(-1)^{\nu}=\prod_{n_{i}=0,1}\prod_{m}\xi_{m}(\bf{\Gamma}_{n_{1}n_{2}}),
\end{align}
where the parity eigenvalues of the occupied bands
at the four time reversal invariant momenta are multiplied.
Because of the time reversal symmetry, each band is doubly degenerate at the time reversal
invariant momentum and every Kramers doublet share the same inversion parity.
The $Z_{2}$ topological invariant counts the parity of one state for each Kramers pair.~\cite{Fu-Kane}

As shown in Table.~\ref{table:z2invariant},
the $Z_{2}$ topological insulators exist when the band filling
is one-quarter or three-quarter.
However, the 3/4 filled case requires unreasonably large $M_{2}$
to achieve an insulating phase.
This is because, as shown in Fig.~\ref{fig:D3M2_kx0}, the overall structures of the band 1 and 2 are in parallel.
To open a full gap
between the upper two bands (band 1 and 2), the magnitude of
the $M_{2}$ density wave order parameter should be as large as their bandwidth.
Therefore the quarter-filled system
is the most favorable for the realization of the $Z_{2}$ topological insulator.

\begin{figure}[t]
\centering
\includegraphics[width=7.5 cm]{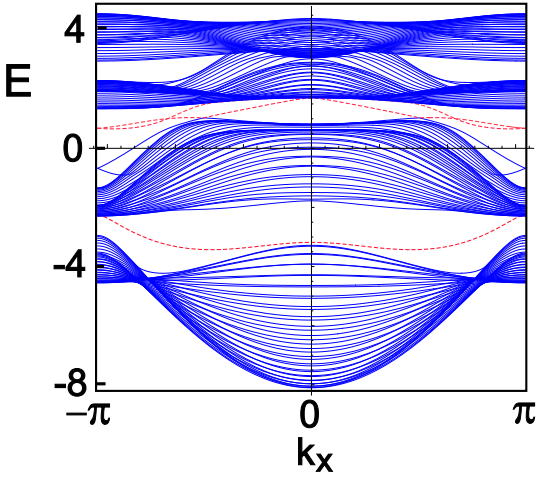}
\caption{(Color online)
Energy spectrum for the topological density wave ground state
with $D_{3}=0.85$ and $M_{2}=0.75$ in a strip geometry.
Here we use open boundary conditions with $N_{y}$=40 sites
along the $y$-direction.
The red dotted lines stand for edge states, all of which are doubly degenerate.
For 1/2-filling, the number of edge states is twice larger than
that for 1/4-filling.
}\label{fig:edgespectrum}
\end{figure}
To support further the nontrivial topological properties of the topological density wave phase with
$D_{3}\neq0$ and $M_{2}\neq0$, we compute the edge state spectrum by
considering the Hamiltonian on a strip geometry, which is infinite
in the $x$-direction but finite in the $y$-direction with open boundaries
at $y$=1 and $y$=$N_{y}$. Here $N_{y}$ indicates the number of lattice sites
in the $y$-direction.
The energy spectrum of the system with $N_{y}=40$ is described in Fig.~\ref{fig:edgespectrum},
which shows
the existence of robust gapless edge states traversing between
the lower two bands (band 3 and 4) and the middle two bands (band 2 and 3).
The upper two bands (band 1 and 2) are not well separated
for $D_{3}=0.85$ and $M_{2}=0.75$, which are used to obtain the energy spectrum.
Each edge state represented by a red dotted line is doubly degenerate,
one with spin-up and the other with spin-down.
For the 1/4-filling with the chemical potential lying in the gap between the band 3 and 4,
there are two gapless edges states on each boundary propagating in
the opposite directions with the opposite spin quantum numbers.
On the other hand, for the 1/2-filling, there are four gapless edge states
on each boundary consistent with the fact that the spin Chern number $C_{S}=4$
whose magnitude is twice larger than that for the 1/4-filling with $C_{S}=-2$.
Therefore for the collinear spin ordering with $M_{2}\neq0$ which
conserves $S_{z}$, there are robust gapless edge states
for both the quarter-filled and half-filled systems.


\subsection{\label{sec:graphene} Comparison to the honeycomb lattice}
\begin{figure}[t]
\centering
\includegraphics[width=8.5 cm]{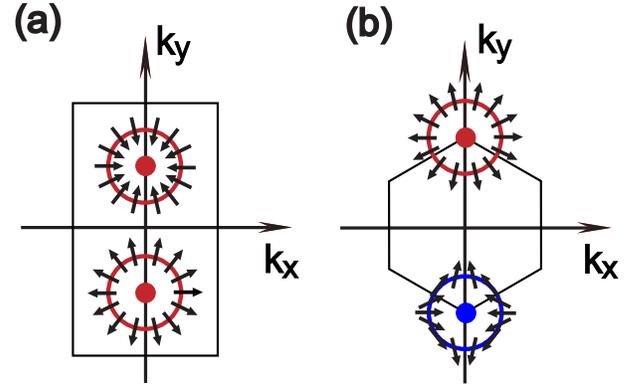}
\caption{(Color online)
The winding directions of the $\vec{d}(\textbf{q})$
vector around the Dirac point. Black arrows
describe the two component $\vec{d}=(d_{x},d_{y})$ vector in Eq.(\ref{eq:dvector})
along the circular path around the Dirac point at the center.
(a) For the two Dirac points between the band 3 and band 4
in the nodal density wave state with the finite $D_{3}$.
The $\vec{d}$ vector has the same winding direction around
the two Dirac points.
(b) For the two Dirac points on the graphene system.
The $\vec{d}$ vector has the opposite winding directions around
the two Dirac points.
}\label{fig:winding}
\end{figure}
It is interesting that a simple on-site density wave order parameter
can generate insulating phases with nontrivial topological properties.
This result can be contrasted with the topological insulator
on the honeycomb lattice where complex second neighbor hopping
processes are required to obtain a topological insulator while the simple
on-site staggered chemical potential gives rise to a trivial band insulator.~\cite{Haldane,KaneMele1,KaneMele2}
It is the distinct topological properties of the Dirac particles in the $D_{3}$ nodal density
wave phase in the $d$-orbital
system, which make it possible to realize the topological insulator
by introducing a simple on-site order parameter $M_{2}$.

In this subsection, we discuss the topological property of
the Dirac particles in the $D_{3}$ nodal density wave state in detail
and compare it with the topological property of the Dirac particles
on the honeycomb lattice.
In the forthcoming discussion we neglect the spin degrees of freedom
and focus on the condition under which the insulating phase
possesses a finite Chern number, which is nothing but a Chern insulator.
Once we find the condition to obtain a Chern insulator, the time reversal invariant topological insulator
can be realized by superposing two Chern insulators with spin-up and spin-down
particles, respectively.

The topological property of the Dirac particles
in the $D_{3}$ nodal density wave state can be understood in the following way.
The low energy Hamiltonian $H^{\text{Dirac}}_{\pm}$ in Eq.(\ref{eq:HDirac2}) for the Dirac particles
near the momentum $\textbf{k}=\textbf{k}_{\pm}$,
where the band touching points between the band 3 and 4 locate, can be written as

\begin{align}\label{eq:HDirac_simple}
H^{\text{Dirac}}_{\pm}(\textbf{q})=&
h_{\pm,x}(\textbf{q})\hat{\tau}_{1}+h_{\pm,y}(\textbf{q})\hat{\tau}_{3}.
\end{align}
Since $\sqrt{h^{2}_{\pm,x}(\textbf{q})+h^{2}_{\pm,y}(\textbf{q})}$ is nonzero away from the degeneracy point,
a two component vector $\vec{d}_{\pm}(\textbf{q})$ with the unit length can be defined as

\begin{align}\label{eq:dvector}
\vec{d}_{\pm}(\textbf{q})=(d_{\pm,x}(\textbf{q}), d_{\pm,y}(\textbf{q}))\equiv&
\frac{(h_{\pm,x}(\textbf{q}), h_{\pm,y}(\textbf{q}))}{\sqrt{h^{2}_{\pm,x}(\textbf{q})+h^{2}_{\pm,y}(\textbf{q})}}.
\end{align}
Along the circle $C_{R}$ satisfying $q^{2}_{x}+q^{2}_{y}=R^{2}\neq0$ with the degeneracy point
at the center, the 2D unit vector $\vec{d}_{\pm}(\textbf{q})$ defines a map
from the circle $C_{R}$ to the unit circle $S^{1}$.
Since the fundamental group $\pi_{1}(S^{1})=Z$,
the 2D unit vector $\vec{d}_{\pm}(\textbf{q})$ has an integer-valued topological
invariant, which is nothing but the winding number $N_{\text{w}}$ defined in Eq.(\ref{eq:winding}).
In terms of the 2D unit vector $\vec{d}_{\pm}(\textbf{q})$, the winding number $N_{\text{w}}$
can be rewritten as,
\begin{align}\label{eq:winding2}
N_{\text{w}} = \frac{1}{2\pi }\oint_{C_{R}} d\theta \hat{z}\cdot (\vec{d}\times \frac{d\vec{d}}{d\theta})
\end{align}
where the loop integral is defined along the circle where the momentum $\textbf{q}=Re^{i\theta}$.~\cite{Volovik}

In Fig.~\ref{fig:winding}(a) we describe the directions
of the two-component $\vec{d}$ vector along the circular path
around the Dirac points for the $D_{3}$ nodal density wave state.
Notice that the $\vec{d}$ vector has the same winding direction with the winding number of $N_{\text{w}}$=1
around the two Dirac points.
The two Dirac points share the same winding number
because of the constraint imposed by lattice symmetries.
Since the first (second) component of
the two-component Dirac field $\Psi_{\mu}(q)$
is given by the $d_{yz}$ ($d_{xz}$) orbital state,
$\Psi_{\mu}(q)$ transforms
to -$\Psi_{\bar{\mu}}(-q)$ under the inversion symmetry due to the
odd parity of $d_{xz}$ and $d_{yz}$ orbitals.
Here $\bar{\mu}$ has the opposite sign of $\mu$.
This imposes the following constraint on the
pair of Dirac Hamiltonians related by the inversion symmetry,
\begin{align}\label{eq:HDirac_constraint}
H^{\text{Dirac}}_{+}(\textbf{q})=&H^{\text{Dirac}}_{-}(-\textbf{q}).
\end{align}
This constraint guarantees the same winding numbers for the two Dirac Hamiltonians
$H^{\text{Dirac}}_{+}$ and $H^{\text{Dirac}}_{-}$.
It is important to notice that every pair of Dirac Hamiltonians
related by the inversion
symmetry satisfies the same constraint
for nodal density wave phases.

The fact that a pair of Dirac Hamiltonians
related by the inversion symmetry have the same winding numbers
is the distinct topological property
of the Dirac particles in the nodal density wave states,
distinguishable from the topological properties of the Dirac particles
in the honeycomb lattice.
In this system, the two Dirac points at the corners of the first Brillouin zone
have the opposite winding directions, which is described in
Fig.~\ref{fig:winding}(b).
Since the inversion symmetry interchanges the two sublattices of the honeycomb lattice,
each of which comprises one component of the Dirac fermion field $\Psi_{\mu}(\textbf{q})$,
the Dirac fermion field $\Psi_{\mu}(\textbf{q})$ transforms, for example,
to $\hat{\tau}_{x}\Psi_{\bar{\mu}}(-\textbf{q})$ under the inversion symmetry.
This imposes the following constraint to the two Dirac Hamiltonians related by the inversion symmetry,
\begin{align}\label{eq:HDirac_constraint2}
H^{\text{Dirac}}_{+}(\textbf{q})=&\hat{\tau}_{x}H^{\text{Dirac}}_{-}(-\textbf{q})\hat{\tau}_{x}.
\end{align}
The additional Pauli matrix reverses the winding direction
of one of the $\vec{d}$ vectors, leading to the two Dirac Hamiltonians
with the opposite winding numbers.
Therefore these two Dirac points can be pair-annihilated when
they are brought together by perturbations.~\cite{Kane_review,Graphene_review}

The relative winding numbers of the pair of the Dirac Hamiltonians related by
the inversion symmetry,
strongly constrain the topological properties of the insulating phases
obtained by mass perturbations to the Dirac particles.
The introduction of a constant mass term $H^{\text{mass}}_{\pm}=m\hat{\tau}_{2}$
to the Dirac Hamiltonian in Eq.~(\ref{eq:HDirac_simple}) gives rises
to the third component $d_{z}(\textbf{q})$ of the corresponding $\vec{d}$ vector.\cite{kaisun1, SpinCh1}
Explicitly, for the massive Dirac Hamiltonian given by
\begin{align}\label{eq:HmassiveDirac_simple}
H^{\text{Dirac}}(\textbf{q})=&
h_{x}(\textbf{q})\hat{\tau}_{1}+h_{y}(\textbf{q})\hat{\tau}_{3}+m\hat{\tau}_{2},
\end{align}
the 3D unit vector $\vec{d}_{3\text{D}}$ is defined as
\begin{align}\label{eq:dvector3d}
\vec{d}_{3\text{D}}(\textbf{q})=&(d_{x}(\textbf{q}), d_{y}(\textbf{q}), d_{z}(\textbf{q}))
\nonumber\\
\equiv&
\frac{(h_{x}(\textbf{q}), h_{y}(\textbf{q}), m)}{\sqrt{h^{2}_{x}(\textbf{q})+h^{2}_{y}(\textbf{q})+m^{2}}}.
\end{align}

If we introduce, for instance, a positive mass term to the two Dirac Hamiltonians
corresponding to the $D_{3}$ nodal density wave phase described in Fig.~\ref{fig:winding}(a),
both of the $\vec{d}_{3\text{D}}$ vectors, which have positive $z$ components,
move along the northern hemisphere as the momentum $\textbf{q}$ sweeps over the two dimensional momentum space.
The net solid angles subtended by these two $\vec{d}_{3\text{D}}$ vectors
over the entire momentum space are the same, each of which covers $2\pi$.
At this point, it is useful to take into account the following relation
between the Chern number of the valence band and the 3D unit vector $\vec{d}_{3\text{D}}$
for the two band Hamiltonian in Eq.~(\ref{eq:HmassiveDirac_simple}),~\cite{kaisun1}
\begin{align}\label{eq:dvector_chern}
N_{C}=&\int\frac{ d^{2}k}{4\pi} \vec{d}_{3\text{D}}\cdot(\partial_{k_{x}}\vec{d}_{3\text{D}}\times\partial_{k_{y}}\vec{d}_{3\text{D}}).
\end{align}
The above identity implies that the Chern number
counts the number of times the 3D unit vector $\vec{d}_{3\text{D}}$ winding around the unit sphere
over the Brillouin zone torus.
Therefore when a constant mass term is added to the two Dirac points
connected by the inversion symmetry, the Chern number $N_{C}=\pm1$
if the two Dirac points have the same winding numbers,
which is realized in the nodal density wave ground state.

In contrast, the net solid angles covered by the two $\vec{d}_{3\text{D}}$ vectors
in the honeycomb lattice have the same magnitudes but with the opposite signs.
Therefore the total solid angle covered by these two $\vec{d}_{3\text{D}}$ vectors vanishes.
The vanishing Chern number contributions from the two Dirac points
leads to the topologically trivial insulating phase
when the simple mass term in Eq.~(\ref{eq:HmassiveDirac_simple}) is introduced.
This contrasting behavior of the pair of 3D unit vectors $\vec{d}_{3\text{D}}$
in these two systems results in the distinct topological
properties of the insulating phases, the topological
insulator in the $d$-orbital system and the topologically trivial
band insulator in the honeycomb lattice when constant mass terms
with the same signs are added to the pair of Dirac points
connected by the inversion symmetry. However, in graphene
if we introduce mass terms with the opposite signs
at the two Dirac points, a topological insulator
with a finite Chern number can be obtained, which
is realized by considering the complex second
nearest neighbor hopping processes on the honeycomb lattice.


\section{\label{sec:phasediagram} Mean field theory}
\begin{figure}[t]
\centering
\includegraphics[width=8.5 cm]{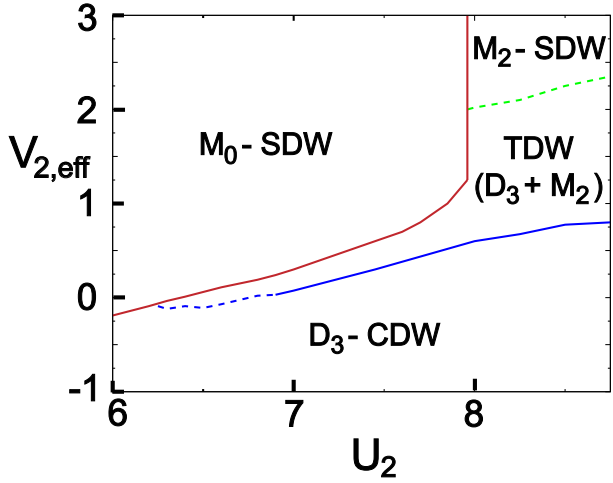}
\caption{(Color online)
Mean field phase diagram. Here we set $U=5$, $J_{H}=0$
and plot the ground state phase diagram varying the inter-orbital on-site repulsion $U_{2}$
and the effective next nearest neighbor repulsion $V_{2,\text{eff}}=V_{2B}-2V_{2A}$.
There is a finite range between $D_{3}$ charge density wave phase ($D_{3}$-CDW)
and $M_{2}$ spin density wave phase ($M_{2}$-SDW)
where the topological density wave phase (TDW) becomes the ground state.
Solid (dotted) lines indicate
the first (second) order phase transitions.
}\label{fig:phasediagram}
\end{figure}

Now we address the question whether the TDW insulator with finite $D_{3}$ and $M_{2}$
can be achieved in real systems.
In particular, by taking into account the interactions between electrons,
we investigate the conditions to realize the TDW insulator via spontaneous symmetry breaking.
Previous mean field studies on the two-orbital Hubbard model
with the hopping Hamiltonian $H_{0}$ in Eq.~(\ref{eq:bareHamiltonian}) show that
the leading instability of the system is the uniform spin density wave phase ($M_{0}$-SDW) which
is described by nonzero $M_{0}$.
However, there exist another density wave order parameters including
imaginary charge and spin density wave states, which
are competing with the uniform spin density wave state ($M_{0}$-SDW) with small energy differences.~\cite{Chubukov,fiveband-Wang,fRG}
Therefore if we include longer range interactions which are not included
in the multi-orbital on-site Hubbard Hamiltonian, another
competing ground states, for example, the topological density wave state (TDW),
can become the ground state replacing the $M_{0}$-SDW state.

Including on-site and inter-site
electron-electron interactions, the full Hamiltonian is given by
\begin{align}
H_{\text{full}}=H_{0}+H_{\text{onsite}}+H_{\text{intersite}}
\end{align}
in which
\begin{align}\label{eq:H_onsite}
H_{\text{onsite}}=&U\sum_{i}\sum_{a=1,2}n_{i,a,\uparrow}n_{i,a,\downarrow}
+U_{2}\sum_{i}n_{i,1}n_{i,2}
\nonumber\\
&+J_{H}\sum_{i}\sum_{\sigma_{1},\sigma_{2}}d^{\dag}_{i,1,\sigma_{1}}d^{\dag}_{i,2,\sigma_{2}}d_{i,1,\sigma_{2}}d_{i,2,\sigma_{1}}
\nonumber\\
&+J_{H}\sum_{i}(d^{\dag}_{i,1,\uparrow}d^{\dag}_{i,1,\downarrow}d_{i,2,\downarrow}d_{i,2,\uparrow}+ \text{H.c.}),
\end{align}
and
\begin{align}\label{eq:H_intersite}
H_{\text{intersite}}=&V_{1A}\sum_{\langle ij \rangle}\sum_{a=1,2}n_{i,a}n_{j,a}
+V_{1B}\sum_{\langle ij \rangle}n_{i,1}n_{j,2}
\nonumber\\
&+V_{2A}\sum_{\langle\langle ij \rangle\rangle}\sum_{a=1,2}n_{i,a}n_{j,a}
+V_{2B}\sum_{\langle\langle ij \rangle\rangle}n_{i,1}n_{j,2},
\end{align}
where $H_{0}$ indicates the hopping Hamiltonian in Eq.(\ref{eq:bareHamiltonian}).
For the on-site interactions described by $H_{\text{onsite}}$, the intra-orbital repulsion $U$,
the inter-orbital repulsion $U_{2}$, and the Hund's coupling $J_{H}$ are considered.
In the $H_{\text{intersite}}$ describing the inter-site Coulomb interactions,
$V_{1A}$ ($V_{1B}$) indicates the nearest-neighbor Coulomb repulsion between
electrons in the same (different) kinds of orbitals. Finally,
$V_{2A}$ ($V_{2B}$) indicates the next nearest-neighbor Coulomb repulsion between
electrons in the same (different) kinds of orbitals.

To investigate the existence of the topological density wave state (TDW) with $D_{3}\neq0$ and
$M_{2}\neq0$ and its competition with the uniform spin density wave phase ($M_{0}$-SDW),
we apply a mean field approximation to the Hamiltonian $H_{\text{full}}$.
The resulting mean-field Hamiltonian is given by
\begin{align}
&H_{\text{MF}}=H_{0}+N\epsilon_{0}
\nonumber\\
&+\sum_{k\in\textbf{RBZ},\sigma}[A_{11,\sigma}\{d^{\dag}_{1,\sigma}(\textbf{k})d_{1,\sigma}(\textbf{k}+\textbf{Q})
+d^{\dag}_{1,\sigma}(\textbf{k}+\textbf{Q})d_{1,\sigma}(\textbf{k})\}
\nonumber\\
&\qquad\quad+A_{22,\sigma}\{d^{\dag}_{2,\sigma}(\textbf{k})d_{2,\sigma}(\textbf{k}+\textbf{Q})
+d^{\dag}_{2,\sigma}(\textbf{k}+\textbf{Q})d_{2,\sigma}(\textbf{k})\}
\nonumber\\
&\qquad\quad+A_{12,\sigma}\{d^{\dag}_{1,\sigma}(\textbf{k})d_{2,\sigma}(\textbf{k}+\textbf{Q})
+d^{\dag}_{1,\sigma}(\textbf{k}+\textbf{Q})d_{2,\sigma}(\textbf{k})\}
\nonumber\\
&\qquad\quad+A_{21,\sigma}\{d^{\dag}_{2,\sigma}(\textbf{k})d_{1,\sigma}(\textbf{k}+\textbf{Q})
+d^{\dag}_{2,\sigma}(\textbf{k}+\textbf{Q})d_{1,\sigma}(\textbf{k})\}],
\end{align}
in which
\begin{align}
\epsilon_{0}=&\frac{U}{8}(M^{2}_{0}-D^{2}_{3})
+\frac{U_{2}}{8}(M^{2}_{2}+2D^{2}_{3})
\nonumber\\
&+\frac{J_{H}}{8}(M^{2}_{0}-D^{2}_{3}-M^{2}_{2})
+(V_{2A}-\frac{V_{2B}}{2})D^{2}_{3},
\end{align}
and
\begin{align}
A_{11,\sigma}=&\frac{U}{4}(D_{3}-\sigma M_{0})-\frac{U_{2}}{2}D_{3}
\nonumber\\
&-(2V_{2A}-V_{2B})D_{3}+\frac{J_{H}}{4}(-D_{3}+\sigma M_{0}),
\nonumber\\
A_{22,\sigma}=&-\frac{U}{4}(D_{3}+\sigma M_{0})+\frac{U_{2}}{2}D_{3}
\nonumber\\
&+(2V_{2A}-V_{2B})D_{3}+\frac{J_{H}}{4}(D_{3}+\sigma M_{0}),
\nonumber\\
A_{12,\sigma}=&A^{*}_{21,\sigma}=\frac{U_{2}}{4}i\sigma M_{2}-\frac{J_{H}}{4}i\sigma M_{2}.
\end{align}
Note that the nearest neighbor Coulomb repulsions $V_{1A}$ and $V_{1B}$
do not contribute to the mean field Hamiltonian
because the order parameters have the ordering wave vector $\textbf{Q}=(\pi,0)$.

The order parameters $M_{0}$, $M_{2}$, $D_{3}$ are determined by solving
the following self-consistent equations,
\begin{align}
M_{0}&=\frac{1}{N}\sum_{\textbf{r}=(r_{x},r_{y})}\sum_{\sigma=\pm}(-1)^{r_{x}}\sigma
\langle d^{\dag}_{\textbf{r},1,\sigma}d_{\textbf{r},1,\sigma}+d^{\dag}_{\textbf{r},2,\sigma}d_{\textbf{r},2,\sigma}\rangle,
\nonumber\\
M_{2}&=\frac{1}{N}\sum_{\textbf{r}=(r_{x},r_{y})}\sum_{\sigma=\pm}(-1)^{r_{x}}\sigma
\langle i d^{\dag}_{\textbf{r},1,\sigma}d_{\textbf{r},2,\sigma}-i d^{\dag}_{\textbf{r},2,\sigma}d_{\textbf{r},1,\sigma}\rangle,
\nonumber\\
D_{3}&=\frac{1}{N}\sum_{\textbf{r}=(r_{x},r_{y})}\sum_{\sigma=\pm}(-1)^{r_{x}}
\langle d^{\dag}_{\textbf{r},1,\sigma}d_{\textbf{r},1,\sigma}-d^{\dag}_{\textbf{r},2,\sigma}d_{\textbf{r},2,\sigma}\rangle.
\end{align}
The chemical potential $\mu$ is also determined self-consistently
to satisfy the half-filling condition.

The resulting mean field phase diagram is shown in Fig.~\ref{fig:phasediagram}.
Here we choose $U=5$, $J_{H}=0$
and compute the ground state phase diagram as a function of the inter-orbital on-site repulsion $U_{2}$
and effective next nearest neighbor repulsion $V_{2,\text{eff}}\equiv V_{2B}-2V_{2A}$.
In the absence of the inter-site interactions $V_{2,\text{eff}}=0$,
the uniform spin density wave phase ($M_{0}$-SDW) dominates the phase
diagram consistent with the previous studies.
However, the inter-orbital on-site repulsion $U_{2}$ suppresses
the uniform spin density wave states ($M_{0}$-SDW) which is diagonal in the orbital space,
but promotes the $M_{2}$-SDW, which is off-diagonal in the orbital space,
to the ground state.
On the other hand, the $D_{3}$ charge density wave phase ($D_{3}$-CDW)
is strongly affected by the next nearest neighbor Coulomb repulsions $V_{2A}$ and $V_{2B}$.
In particular, the $V_{2A}$, the next nearest neighbor repulsion between the electrons
in the same kinds of orbitals, strongly favors the $D_{3}$-CDW
because the staggered orbital ordering described by $D_{3}$-CDW can avoid
the energy cost coming from $V_{2A}$.
Notice that there is a finite range in the parameter space where
both $M_{2}$-SDW and $D_{3}$-CDW are nonzero realizing
the topological density wave phase (TDW).

The above mean field phase diagram is obtained for the half filled case
where the topological property of the TDW phase is not robust
against perturbations breaking $S_z$ symmetry.
On the other hand, the TDW insulator at 1/4 filling maintains its topological properties
as long as time reversal symmetry is preserved.
It was shown, in the study of the single orbital extended Hubbard model,
that the next nearest neighbor interaction ($V_{2A}$ for our model) stabilizes
a stripe pattern charge ordering with the momentum ${\bf Q}=(\pi,0)$.~\cite{ogata}
This occurs at 1/4 filling when the on-site Hubbard interaction $U$
is much stronger than the hopping amplitude $t$ satisfying $U >> t$, so that double occupancy is almost frozen.
In our model,  there are two orbitals of $d_{xz}$ and $d_{yz}$.  Similar to the single orbital case,
we find that the next nearest neighbor interaction stabilizes $D_3$ order.
Therefore, when the on-site intra- and inter-orbital
interactions satisfy $U, U_2 \gg t$,  the $D_{3}$-CDW ordering should be favored at 1/4 filling.

To get a TDW insulator, finite $M_2$ is required in addition to $D_3$.
As shown in Sec. II B, the $M_2$ order order parameter is equivalent to a staggered spin-orbit coupling,
$ \sim\sum_{i}(-1)^{i_x} S_{i,z} L_{i,z}$.
One can show that when spin-orbit interaction is present, $M_2$ term can be induced as long as $D_3$ sets in,
since $D_3$ leads to unequal density between $d_{xz}$ and $d_{yz}$ orbitals.
Therefore, we expect that the TDW insulator can be
obtained by tuning $V_{2B}$ when $U, U_2 \gg t$ at 1/4 filling in the presence of the spin-orbit coupling.

\section{\label{sec:threeorbital} Topological insulators in three-band systems}

In the preceding sections, we have focused on a two-band tight-binding
Hamiltonian, which consists of $d_{xz}$ and $d_{yz}$ orbitals.
However, the main idea for realizing topological insulators using two density
wave order parameters with opposite symmetries under reflections is valid in general
and applicable to more realistic multi-orbital systems.
Here we extend our analysis to a three-band model composed of
$d_{xz}$, $d_{yz}$, and $d_{xy}$ orbitals.
In particular, we apply our idea to a more realistic model Hamiltonian
relevant to the iron pnictide system,
which is a representative itinerant multi-band system
manifesting a density wave ground state with the ordering
wave vector $\textbf{Q}=(\pi,0)$.~\cite{threeband-Dagotto,threeband-LeeWen,fiveband-Graser,fiveband-Kuroki,fiveband-Wang}
Here we adopt the three-orbital model proposed by Daghofer et al.,\cite{threeband-Dagotto}
which captures the main physical properties of the Fe-pnictide systems.
In the momentum space, the effective three-band tight-binding Hamiltonian
is given by

\begin{align}\label{eq:threeband_Hamiltonian}
\text{H}_{\text{3band}}=\sum_{\textbf{k},\sigma}\sum_{\mu,\nu} d^{\dag}_{\mu,\sigma}(\textbf{k})
T^{\mu\nu}(\textbf{k})
d_{\nu,\sigma}(\textbf{k}),
\end{align}
where
\begin{align}\label{eq:threeband_parameters}
T^{11}&=2t_{2}\cos k_{x} + 2t_{1}\cos k_{y} +4t_{3}\cos k_{x}\cos k_{y} - \mu_{c},
\nonumber\\
T^{22}&=2t_{1}\cos k_{x} + 2t_{2}\cos k_{y} +4t_{3}\cos k_{x}\cos k_{y} - \mu_{c},
\nonumber\\
T^{33}&=2t_{5}(\cos k_{x} + \cos k_{y}) + 4t_{6}\cos k_{x}\cos k_{y} - \mu_{c} + \Delta_{xy},
\nonumber\\
T^{12}&=T^{21}=4t_{4}\sin k_{x}\sin k_{y},
\nonumber\\
T^{13}&=(T^{31})^{*}=2it_{7}\sin k_{x} + 4it_{8}\sin k_{x}\cos k_{y},
\nonumber\\
T^{23}&=(T^{32})^{*}=2it_{7}\sin k_{y} + 4it_{8}\sin k_{y}\cos k_{x}.
\end{align}

Here we use the unfolded Brillouin zone satisfying $-\pi<k_{x}, k_{y}\leq \pi$
as before, which corresponds to one iron atom per unit cell.
In real iron pnictide materials, the unit cell
contains two Fe atoms due to the buckling of the As atoms. Therefore
the unit translations along the $x$ ($T_{x}$) and $y$ ($T_{y}$) directions by the nearest neighbor Fe-Fe distance,
are not the symmetries of the system.
However, as pointed out in Ref.~\onlinecite{threeband-LeeWen},
the system is invariant under the translations combined with the reflection $P_{z}$ with respect to the $xy$ plane,
i.e. $P_{z}T_{x}$ and $P_{z}T_{y}$.
Then the eigenstates can be labeled by a pseudo-crystal momentum corresponding
to the eigenvalues of the combined operations $P_{z}T_{x}$ and $P_{z}T_{y}$
with one iron atom per unit cell.
We use this pseudo-crystal momentum to label states
for the momentum space representation of the Hamiltonian in Eq.~(\ref{eq:threeband_Hamiltonian}).

In Eq.~(\ref{eq:threeband_Hamiltonian}) and (\ref{eq:threeband_parameters}),
$\mu$=1, 2, 3 indicate $d_{xz}$, $d_{yz}$, and $d_{xy}$ orbitals, respectively.
$\Delta_{xy}$ represents the atomic potential of $d_{xy}$ orbital
relative to $d_{xz}$ and $d_{yz}$ orbitals.
The chemical potential is given by $\mu_{c}$.
The hopping parameters are displayed in Table~\ref{table:3band-hoppingparameters},
which are determined in Ref.~\onlinecite{threeband-Dagotto}.
For the 2/3 filling,\cite{threeband-Dagotto} there are two hole pockets near the $\Gamma$ point
and two electron pockets at the $X$ and $Y$ points, which are consistent
with the LDA calculations and ARPES measurement for LaOFeAs.

\begin{table}
\begin{tabular}{@{}|ccccccccc|}
\hline \hline
$t_{1}$& $t_{2}$ & $t_{3}$ & $t_{4}$ & $t_{5}$  & $t_{6}$ & $t_{7}$ & $t_{8}$ & $\Delta_{xy}$ \\
\hline
0.02 & 0.06 & 0.03 & -0.01 & 0.2 & 0.3 & -0.2 & 0.12 & 0.4\\
\hline \hline
\end{tabular}
\caption{Parameters for the three-band tightbinding Hamiltonian.~\cite{Footnote1}}
\label{table:3band-hoppingparameters}
\end{table}

Density wave order parameters with the momentum $\textbf{Q}$=($\pi$,0) can be described
by the following Hamiltonian,

\begin{align}
\hat{H}_{\text{CDW}}&=\sum_{i,\sigma}\sum_{a,b=1}^{3}(-1)^{i_{x}}D_{ab}
d^{\dag}_{i,a,\sigma}d_{i,b,\sigma},
\nonumber\\
\hat{H}_{\text{SDW}}&=\sum_{i,\sigma_{1}\sigma_{2}}\sum_{a,b=1}^{3}(-1)^{i_{x}}M_{ab}
d^{\dag}_{i,a,\sigma_{1}}s^{z}_{\sigma_{1}\sigma_{2}}d_{i,b,\sigma_{2}}.
\end{align}
The order parameter represented by 3 $\times$ 3 Hermitian matrix $\hat{D}$ ($\hat{M}$) has nine independent components $D_{ij}$ ($M_{ij}$).
The transformation properties of these density wave order parameters
under the reflections $P_{x}P_{z}$ and
$P_{y}P_{z}$, inversion $I$, and time-reversal $T$
are summarized in Table~\ref{table:threeband_CDW} and ~\ref{table:threeband_SDW}.
Notice that the reflections $P_{x}$ and $P_{y}$
are not the symmetries of the system.
The Hamiltonian is invariant only under the combined transformations
$P_{x}P_{z}$ and $P_{y}P_{z}$.

\begin{table}
\begin{tabular}{@{}|c|c|c|c|c|c|c|c|c|c|}
\hline \hline
& $D_{11}$ & $D_{22}$ & $D_{33}$ & $D^{R}_{12}$  & $D^{I}_{12}$ & $D^{R}_{13}$ & $D^{I}_{13}$ & $D^{R}_{23}$ & $D^{I}_{23}$\\
\hline \hline
$P_{x}P_{z}$ & + & + & + & - & - & - & - & +  & + \\
$P_{y}P_{z}$ & + & + & + & - & - & + & + & -  & - \\
$I$          & + & + & + & + & + & - & - & -  & - \\
$T$          & + & + & + & + & - & + & - & +  & - \\
\hline \hline
\end{tabular}
\caption{Symmetry of the charge density order parameters
with the momentum ($\pi$,0). Here `+' (`-')
indicates `even' (`odd') parity of order parameters
under the corresponding symmetry operation.
The complex off-diagonal components $D_{ij}$ ($i\neq j$) are decomposed as
$D_{ij}=D^{R}_{ij}+iD^{I}_{ij}$.}
\label{table:threeband_CDW}
\end{table}
\begin{table}
\begin{tabular}{@{}|c|c|c|c|c|c|c|c|c|c|}
\hline \hline
& $M_{11}$ & $M_{22}$ & $M_{33}$ & $M^{R}_{12}$  & $M^{I}_{12}$ & $M^{R}_{13}$ & $M^{I}_{13}$ & $M^{R}_{23}$ & $M^{I}_{23}$\\
\hline \hline
$P_{x}P_{z}$ & + & + & + & - & - & - & - & +  & + \\
$P_{y}P_{z}$ & + & + & + & - & - & + & + & -  & - \\
$I$          & + & + & + & + & + & - & - & -  & - \\
$T$          & - & - & - & - & + & - & + & -  & + \\
\hline \hline
\end{tabular}
\caption{Symmetry of the spin density order parameters
with the momentum ($\pi$,0). Here `+' (`-')
indicates `even' (`odd') parity of order parameters
under the corresponding symmetry operation.
The complex off-diagonal components $M_{ij}$ ($i\neq j$) are decomposed as
$M_{ij}=M^{R}_{ij}+iM^{I}_{ij}$.}
\label{table:threeband_SDW}
\end{table}

To obtain a nodal spin density wave ground state,
we consider the simplest uniform charge density wave order parameter,
$\hat{D}_{\text{uniform}}\equiv$diag[$d_{0}$,$d_{0}$,$d_{0}$]
with the finite diagonal components of $D_{11}=D_{22}=D_{33}=d_{0}$.
This generates many Dirac points along an axis with a reflection symmetry
in the momentum space, whenever a band touching
occurs between two bands with opposite reflection parities.
Now let us introduce another density wave order parameter
to get a gapped phase. To open a full gap between neighboring
bands we need a density wave order parameter which is odd
under the reflection symmetries $P_{x}P_{z}$ and $P_{y}P_{z}$.
Imposing the time reversal symmetry, the imaginary
part of the spin density wave order parameter $M^{I}_{12}\equiv \text{Im}(M_{12})$,
is the unique choice to obtain a topological insulator.
\begin{figure}[t]
\centering
\includegraphics[width=7 cm]{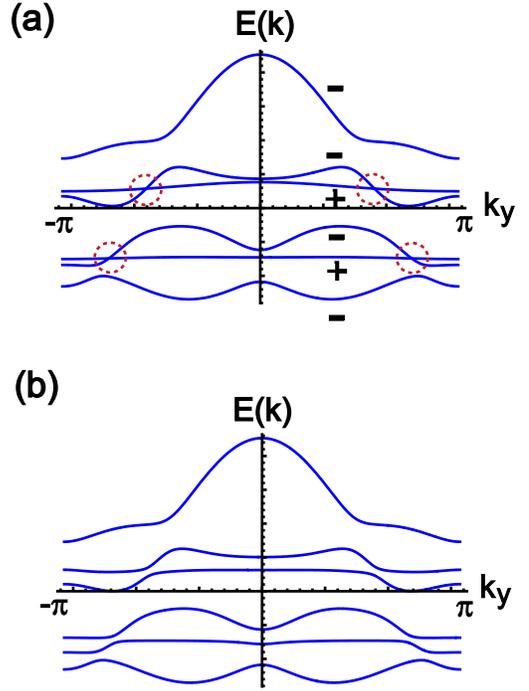}
\caption{(Color online)
The band structures of the density wave states
with the momentum $\textbf{k}$=($\pi$,0) along $k_{y}$ axis.
(a) For the uniform charge density wave state with $D_{\text{uniform}}=0.5$.
The locations of Dirac nodes are indicated by red dotted circles.
The parities of the bands under the $P_{x}P_{z}$ symmetry are indicated by + (even) and - (odd).
(b) A density wave phase with $D_{\text{uniform}}\neq0$ and $M^{I}_{12}\neq0$ at
the same time. Here we take $D_{\text{uniform}}$=0.5 and $M^{I}_{12}$=0.1.
Each band is well-separated from the other bands.
} \label{fig:threeband_nodes}
\end{figure}

In Fig.~\ref{fig:threeband_nodes} (a) we plot the band structure
of the uniform charge density wave state with nonzero $D_{\text{uniform}}\neq0$
along the $k_{y}$ axis. Since $D_{\text{uniform}}$
preserves the $P_{x}P_{z}$ reflection symmetry,
each band has a definite reflection parity under $P_{x}P_{z}$.
The reflection parities of the bands are also indicated in Fig.~\ref{fig:threeband_nodes} (a).
Notice that nodal points exist between a pair of the bands with opposite
reflection parities, which are indicated
by red dotted circles in Fig.~\ref{fig:threeband_nodes} (a). However, once we introduce a nonzero $M^{I}_{12}$,
these nodal points disappear and a fully gapped phase with well-separated bands emerges.
The band structure of the
resulting gapped phase is described in Fig.~\ref{fig:threeband_nodes} (b).

\begin{table}
\begin{tabular}{@{}|c|c|c|c|}
\hline \hline
& $M^{I}_{12}=0.1$ & $M^{I}_{12}=0.2$ & $M^{I}_{12}=0.3$ \\
\hline \hline
$C_{S,1}$ & 0 & -2 & -2 \\
$C_{S,2}$ & 0 & +2 & +2 \\
$C_{S,3}$ & 0 & 0 & -4 \\
$C_{S,4}$ & -2 & -2 & +2 \\
$C_{S,5}$ & +2 & +2 & +2 \\
$C_{S,6}$ & 0 & 0 & 0 \\
\hline \hline
\end{tabular}
\caption{The spin Chern numbers of the bands for
a gapped density wave phase
with $D_{\text{uniform}}\neq0$ and $M^{I}_{12}\neq0$.
Here we set $D_{\text{uniform}}=0.5$ and change
the magnitude of $M^{I}_{12}$.
}
\label{table:Chernnumbers_threeband}
\end{table}

To investigate the topological property of the
gapped phase, we compute the spin Chern numbers of the bands.
Since the $z$ component of the spin $S_{z}$ is still conserved,
the spin Chern number is a well-defined quantity.
In Table~\ref{table:Chernnumbers_threeband},
we show the distribution of the spin Chern numbers for several values of $M^{I}_{12}$
supporting fully gapped phases.
Here $C_{S,n}$ indicates the spin Chern number of
the $n$th band defined as
$C_{S,n}=N_{C,n,\uparrow}-N_{C,n,\downarrow}$
where $N_{C,n,\uparrow}$(=-$N_{C,n,\downarrow}$) denotes the Chern number of
the $n$th spin-up band.
We label that the band 1 has the highest energy
and the band index $n$ increases as the energy eigenvalue decreases.
In the case of the gapped phase that is obtained by adding a small $M^{I}_{12}$
on the nodal charge density wave state with $D_{\text{uniform}}\neq0$,
only the 4th and 5th band support nonzero spin Chern numbers
shown in the 2nd column of Table~\ref{table:Chernnumbers_threeband}.
Interestingly, however, as the magnitude of $M^{I}_{12}$ increases,
band gap closing and reopening occur successively.
For instance, for the uniform density wave state with $D_{\text{uniform}}=0.5$,
the first gap closing happens between the band 1 and 2 for $M^{I}_{12}\approx0.15$.
As $M^{I}_{12}$ increases further, another fully gapped phase is obtained
with the spin Chern numbers displayed in the 3rd column of Table~\ref{table:Chernnumbers_threeband}.
It is interesting to notice that after the gap closing and reopening process,
the number of the bands supporting finite spin Chern numbers has increased.
Similar gap closing happens again for $M^{I}_{12}\approx0.22$
leading to the redistribution of the spin Chern numbers shown in the last column
of Table~\ref{table:Chernnumbers_threeband}.
Note that all cases with 1/3 filling give TDW insulators, while TDW phase with 5/6 filling
occurs only for the $M^{I}_{12} =0.2 $ and $0.3$.

\section{\label{sec:conclusion} Summary and  Discussion }

In this paper, we investigate theoretically
if topological insulators can be achieved from
a nodal density wave state with broken translational symmetry.
While a nonzero density wave order parameter in general opens a gap between the degenerate states connected by
the ordering wave vector, nodal density wave phases occur in multi-orbital systems via translational symmetry breaking
due to the distinct symmetry properties of orbitals.
Such a nodal density wave state supports a large number of Dirac nodes between neighboring bands.
We have explicitly proved that a pair of inversion symmetric Dirac points
share the same topological winding numbers in nodal density wave states
contrary to the Dirac points in the honeycomb lattice.
If we introduce an additional order parameter
whose transformation property under reflection symmetries is
opposite to that of the underlying order parameter,
the system can be a gapped insulator at certain filling factors.
Among those insulators, time-reversal invariant TDW insulators with helical edge states
are identified.

The existence of a nodal density wave
ground state is experimentally verified in
a recent ARPES measurement on BaFe$_{2}$As$_{2}$~\cite{ARPES}
and quantum oscillation experiments on BaFe$_{2}$As$_{2}$ and SrFe$_{2}$As$_{2}$.~\cite{QO1,QO2,QO3}
It is interesting to notice that, according to these experimental studies,
the velocity of Dirac fermions is estimated to be 14 - 20 times slower
than that in graphene.~\cite{QO3} This implies that the Dirac fermions in nodal density wave states
are more susceptible to interaction effects.
However, according to our mean field calculation,
it seems to be difficult to realize quantum spin Hall insulators in Fe pnictides system,
as it favors a conventional spin density wave state ($M_0$).

Our results in general imply that transition metal materials with
two-dimensional square lattice structure possessing
partially filled $t_{2g}$ orbitals
are good candidates for TDW insulators.
In particular,
in the case of the effective two-orbital (three-orbital) model,
the 1/4 filled (1/3 filled) system is the most promising for the realization of
TDW insulators.
However, to make a prediction on real materials with layered perovskite structure, it is important to generalize
our study to three dimensional systems taking into account interlayer couplings.
Stacking of two dimensional TDW insulators
simply leads to a weak topological insulator.~\cite{Fu-Kane-Mele}
Therefore identifying three dimensional TDW phases with a nontrivial strong topological invariant
in the layered perovskite structure is an interesting but challenging future work.

Finally,
it is  worthwhile to comment the consequences of relaxing the constraint of time-reversal invariance.
When an imaginary charge density wave state ($D_2$) breaking time-reversal symmetry
occurs in the presence of a nodal density wave state,
a gapped topological phase with topologically protected edge modes can be developed.
In contrast to the case of the quantum spin Hall insulator,
here the spin-up and down bands have the same Chern number,
which gives rise to an insulator with finite Hall conductance.
Interestingly, the imaginary charge density wave state
is one of the competing ground states in iron pnictide systems,~\cite{podolsky,Chubukov,fRG}
which is expected to be achieved in real materials.~\cite{ODW}
Thus searching for gapped phases
proximate to nodal density wave states is a new avenue to topological phases.

\acknowledgments

We thank Daniel Podolsky for helpful discussions.
This work was supported by the NSERC of Canada, the Canada Research Chair, and
the Canadian Institute for Advanced Research.






\begin{thebibliography}{99}
\bibitem{Footnote1}
In Ref.\onlinecite{threeband-Dagotto}, $t_{8}$ and $t_{7}$
was fixed by $t_{8}=-t_{7}/2$ to make
the electron pockets have pure $xy$ character along $\Gamma-X$
and $\Gamma-Y$ directions.
Since this additional constraint generates accidental band crossing points,
we introduce 10 percent modulation in the magnitude of $t_{8}$, which does not
affect the overall band structure and the shape of the Fermi surfaces.
\end{thebibliography}
\end{document}